\documentstyle[aps,prc,epsf,tighten]{revtex}
\renewcommand{\Re}{{\mathrm{Re}}}

\newcommand{\dida}[1]{/ \!\!\! #1}
\newcommand{\didag}[1]{/ \!\!\!\!\!\, #1}

\newcommand{\abb}{2m(\nu-\nu_B)}
\newcommand{\abbb}{m(\nu-\nu_B)}
\newcommand{\abc}{(m+m_\Delta)}
\newcommand{\abd}{\frac{2}{3m_\Delta^2}}
\newcommand{\abe}{\frac{1}{3m_\Delta}}
\newcommand{\abf}{(m_\Delta-m)}
\newcommand{\abg}{\frac{q^2}{2m\nu_B}}
\newcommand{\dotl}{..........}
\newcommand{\dashl}{- - - - -}
\newcommand{\ldashl}{-- -- -- --}
\newcommand{\solidl}{---------}
\newcommand{\dashdotl}{\_.\_.\_.\_.\_}

\newcommand{\gae}{g_{\mbox{\tiny{\sl A}}}^e}
\newcommand{\gve}{g_{\mbox{\tiny{\sl V}}}^e}

\tighten
\begin{document}
\preprint{MKPH-T-95-10}
%\draft
\title{\hfill MKPH-T-95-10 \\ \ \\ \ \\
Parity violating pion electroproduction off the nucleon
\thanks{This work has been supported by the Deutsche
For\-schungs\-ge\-mein\-schaft (SFB 201).}}
\author{H.-W. Hammer and D. Drechsel}
\address{Institut f\"ur Kernphysik, Johannes
Gutenberg-Universit\"at Mainz, D-55099 Mainz, Germany}
%\date{\today}
\maketitle
%%%%
\begin{abstract}
Parity violating (PV) contributions due to interference between
$\gamma$ and $Z^0$ exchange are calculated for pion
electroproduction off the nucleon.
A phenomenological model with effective Lagrangians is
used to determine the resulting asymmetry for the energy region
between threshold and $\Delta(1232)$ resonance. The $\Delta$ resonance
is treated as a Rarita-Schwinger field with phenomenological
$N \Delta$ transition currents. The background contributions are given
by the usual Born terms using the pseudovector $\pi N$ Lagrangian.
Numerical results for the asymmetry are presented.
\\ \\
PACS numbers: 25.30.Rw, 13.10.+q
\end{abstract}
%%%%
\section{Introduction}
\label{int}
The study of the electroweak interaction between leptons and
hadrons has been a challenging topic ever since the standard model was
proposed by Glashow, Salam and Weinberg. This model predicts the coupling
of the electroweak currents to leptons and quarks in terms of the electric
charge $e$ and the Weinberg angle $\theta_{W}$. In particular, the weak
neutral current is mediated by the exchange of the $Z^0$ gauge boson. At
low and moderate momentum transfer its contribution is suppressed relative
to photon exchange by a factor $q^2/M_Z^2$, where $q$ is the four-momentum
of the exchanged boson and $M_Z$ the mass of the $Z^0$. The interference
term beween photon and $Z^0$ exchange contains a parity violating (PV)
effect, which becomes visible as an asymmetry by scattering polarized
electrons with helicity along the direction of the beam $h = +1$ or
opposite to it $h = -1$,
\begin{equation}
\label{def_a}
A = \frac{ \sigma(h = +1) - \sigma(h = -1)}{ \sigma(h = +1)
+ \sigma(h = -1)} \, .
\end{equation}
The quantity $\sigma$ in this equation should represent an inclusive
cross section. In the case of a coincidence experiment, e.g.
$e + N \rightarrow e + N + \pi$,
there also appear parity conserving asymmetries due to the
electromagnetic interaction. Specifically, the so-called 5th response
function will generate a background of helicity-dependent contributions,
which are parity conserving and, therefore, generally larger than
the PV effects by several orders of magnitude.

The pioneering experiment to measure PV asymmetries has been performed at
SLAC \cite{Pr78} by deep inelastic electron scattering
on a deuterium target. This
experiment has been followed by investigations of PV quasifree scattering
off $^9 B$ at Mainz \cite{He89}
and PV elastic scattering off $^{12} C$ at MIT/Bates \cite{So90}.
In the latter case the momentum transfer was only
$q^2 = - (150 \mbox{ MeV})^2$,
leading to the tiny asymmetry
$A = (0.60\pm 0.14 \pm 0.02) \cdot 10^{-6}$.
These experiments were originally devised to test the
standard model, in particular to measure the Weinberg angle. The value
obtained for this angle by high-energy experiments was confirmed within
the error bars of about 10\%. With the advent of new electron accelerators
like CEBAF, MAMI and MIT/Bates, having high intensity, high duty-factor
and
a polarized beam, the quality of the data can be considerably improved.
Moreover, with $\sin^2\theta_W = 0.2319 (5)$ known to 4 decimal
places and the
standard model firmly established, the strategy of the new experiments will
be redirected towards an improved understanding of the structure of the
nucleon. In particular, PV elastic electron scattering will provide
information on the strangeness content of the nucleon. Three such
experiments are being planned with the 4 GeV CEBAF beam
\cite{Be92,So91,Bei91} and there are also
proposals to measure the electric radius and the magnetic moment of
strange quark pairs at MAMI \cite{Ha93}
and MIT/Bates \cite{MK89} in the region of 1 GeV.

Assuming that the present experimental activities will soon yield novel
information on the ground state of the nucleon, we deem it appropriate to
study the effect of PV interactions for inelastic processes. It is
therefore the aim of this contribution to investigate the PV asymmetries
for electroproduction of pions. Previous calculations of
Nath et al.\cite{Na82} and Jones et al.\cite{Jo80}
at medium energies and of Cahn et al.\cite{Ca78} at higher
energies have been based on the production of stable $\Delta$ isobars.
However, there should be non-negligible background contributions
interfering coherently with the resonance. Estimates for such
contributions
have been reported earlier by Ishankuliev et al.\cite{Is80}
and by Li et al.\cite{Li82}. In particular, Li et al.\cite{Li82}
have also considered PV effects in the hadronic
wave function, i.e. at the $\pi N$ vertex. It turns out,
however, that such
contributions are of the order of $10^{-7}$, much smaller than the
expected asymmetries due to PV interferences in the electroweak
interaction, which are of the order of $10^{-4} q^2/\mbox{ GeV}^2$.

In the following section we will briefly review the kinematics, the cross
section and the decomposition of both vector and axial currents into the
invariant amplitudes. Sect. 3 presents our model using effective
Lagrangians. It includes a background of Born terms with pseudovector
$\pi N$ coupling and the $\Delta$ isobar treated as a Rarita-Schwinger
field with phenomenological $N\Delta$ transition currents.
Assuming that the hadronic currents are dominated
by $u$ and $d$ quarks, the weak neutral current may
be decomposed in terms of strong isospin \cite{Mu94}
and related to hadronic currents.
The matrix elements are then decomposed into invariant
amplitudes \cite{Ad68}
and according to their isospin structure. The numerical results are
presented in Sect. 4. As a test we have first calculated the inclusive
electromagnetic cross section.
The experimental data for this process \cite{Ly67,Co67}
can be well reproduced by the model. We present our results for PV
asymmetries as function of excitation energy, scattering angle and
momentum and compare our asymmetries with previous calculations. Finally,
we give a summary and some conclusions in Sect. 5.
%%%%
\section{Formalism}
\label{form}
%%%%
\subsection{Kinematics}
\begin{figure}[htb]
\centerline{
\epsfxsize=8.0cm
\epsffile{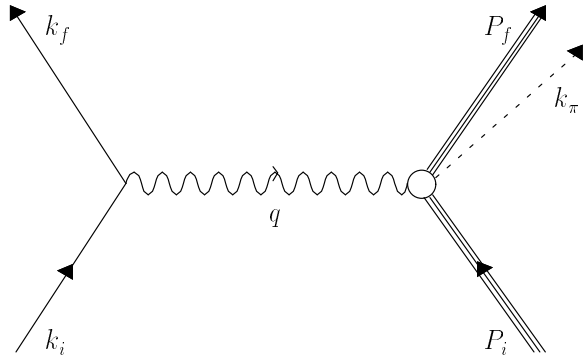}}
\caption{\label{fig1}
Kinematic variables for pion electroproduction
(see text for notation).}
\end{figure}

\begin{figure}[htb]
\centerline{
\epsfxsize=8.0cm
\epsffile{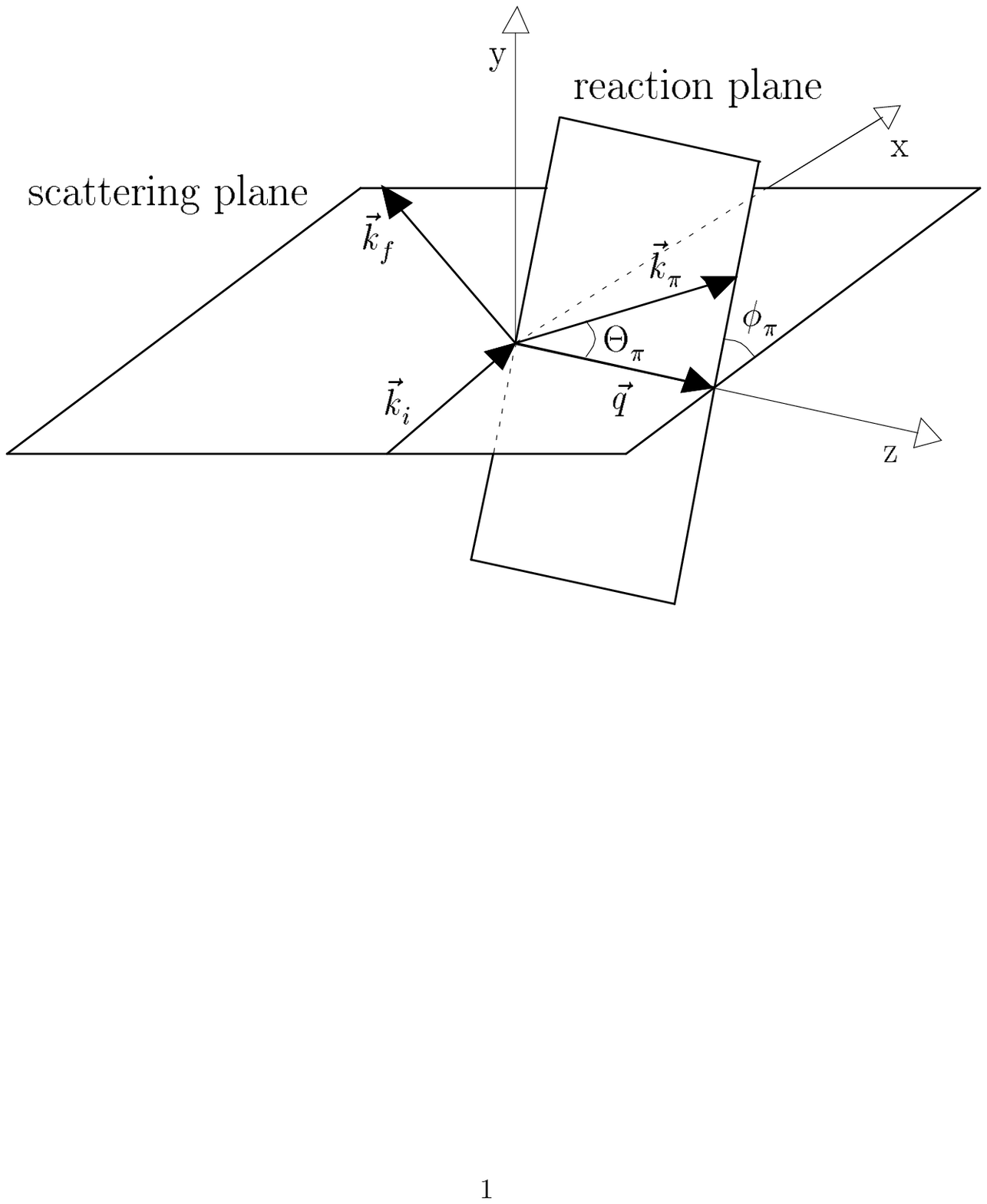}}
\caption{\label{fig2}
Kinematics and angles for pion electroproduction
(see text for notation).}
\end{figure}

We consider the reaction shown in Fig. \ref{fig1} .
$P_i = (E_i,\vec{P_i})$ and $P_f = (E_f,\vec{P_f})$ are the
4-momenta of the nucleon in the initial and the final state,
respectively, the produced pion has momentum
$k_\pi = ( \omega_\pi,\vec{k_\pi})$. The momentum transfer
$q = ( \omega,\vec{q} )$ is the difference of the 4-momenta
of the ingoing and outgoing electron $k_i - k_f$, with $k_l =
( \varepsilon_l,\vec{k_l})$. The spins of nucleons and electrons
in the initial and final states are denoted by $S_i$, $S_f$, $s_i$,
and $s_f$. If not stated otherwise, the kinematical
variables are evaluated in the
laboratory system, which is defined by $P_i = (m,\vec {0})$.
Furthermore, the Mandelstam variables
\begin{eqnarray}
s &=& ( P_i + q )^2 ,\quad
t = ( q - k_\pi)^2 , \quad
u = ( P_i - k_\pi)^2 ,
\end{eqnarray}
are equivalent to the following set of Lorentz invariant
kinematic variables \cite{Ad68}:
\begin{eqnarray}
\nu &=& \frac{P \cdot q}{m} \, ,\qquad
\nu_B = -\frac{ k_\pi \cdot q }{ 2 m} \, , \qquad
W = \sqrt{s} \, ,
\end{eqnarray}
with $P = \frac{1}{2}(P_i + P_f)$.
The latter set of variables will be used from now on.
Besides, the coincidence experiment is characterized by the angles
shown in Fig. \ref{fig2} . The polar angle
$\Theta_\pi$ is the angle between the pion 3-momentum $\vec{k}_\pi$
and the momentum transfer $\vec {q}$, the azimuthal angle
$\phi_\pi$ is the angle between the scattering plane,
defined by $\vec{k_i}$ and $\vec{k_f}$,
and the reaction plane, spanned by $\vec{k_\pi}$
and $\vec{q}$. Furthermore, the scattering angle of
the electron is $\Theta_e$. In the following sections we will
be interested in the inclusive cross section, i.e. the
cross section has to be integrated over the pion angles.
%%%%
\subsection{Invariant matrix element}
The differential cross section \cite{Re90}
\begin{equation}
\label{wq}
d \sigma =
\frac{ (2 \pi)^4 \delta^4( P_i + q - P_f - k_\pi )}{4 [ (P_i
\cdot k_i)^2 - m_e^2 m^2 ]^{\frac{1}{2}}}
\prod^{n_f}_{j = 1} \frac{ d^3 \vec{p}_j }{ ( 2 \pi)^3 2 E_j }
\mid {\cal M}_{fi} \mid^2 ,
\end{equation}
has to be integrated over the momenta of the pion and
the final nucleon.
In the one boson exchange approximation and for
momentum transfers $q^2 \ll {M_Z}^2$,
the invariant matrix element is of the form
\begin{eqnarray}
\frac{q^4}{e^4} \mid {\cal M}_{fi} \mid^2 &=&
\mid  j_\nu^{EM} J^\nu_{EM}
+ \frac{q^2}{{M_Z}^2} j_\nu^{NC} J^\nu_{NC} \mid^2
+ O( \frac{q^4}{{M_Z}^4} ) \\
& &=  {j_\mu^{EM}}^\dagger j_\nu^{EM} {J^\mu_{EM}}^\dagger J^\nu_{EM}
+  \frac{q^2}{{M_Z}^2} \left( {j_\mu^{EM}}^\dagger j_\nu^{NC}
{J^\mu_{EM}}^\dagger J^\nu_{NC} + h.c. \right)
+ O( \frac{q^4}{{M_Z}^4} ) \nonumber \\
& &= \eta_{\mu\nu}^{EM} W^{\mu\nu}_{EM}
+ \frac{q^2}{{M_Z}^2} \eta_{\mu\nu}^{I} W^{\mu\nu}_{I}
+ O( \frac{q^4}{{M_Z}^4} ) \; . \nonumber
\end{eqnarray}
In this expression, the photon couples to the electromagnetic
current of the nucleon, $J_\nu^{EM} = V_\nu^{EM}$, and the $Z^0$
gauge boson to the familiar combination of vector and axial currents
$J_\nu^{NC} = V_\nu^{NC} + A_\nu^{NC}$. The corresponding
currents of the electron are denoted by $j_\nu^{EM}$ and
$j_\nu^{NC}$. These currents may be combined to
$ W^{\mu\nu}$ and $\eta_{\mu\nu}$, the hadronic and leptonic
tensors. The tensors may be decomposed into symmetric and
antisymmetric parts. If we neglect the electron mass, the
leptonic tensor is
\begin{eqnarray}
\label{lept}
\eta_{\mu\nu}^{(s)} &=& 2 ( 2 K_\mu K_\nu + \frac{1}{2}
( q^2 g_{\mu\nu} - q_\mu q_\nu ) ) \, ,\\
\eta_{\mu\nu}^{(a)} &=&
- 2 i \epsilon_{\mu\nu\alpha\beta} q^\alpha K^\beta \, , \nonumber
\end{eqnarray}
with $q = ( k_i - k_f )$ and $K = \frac{1}{2} ( k_i + k_f )$.
A reasonable definition for the hadronic tensor is
\begin{eqnarray}
\label{w}
W_{\mu\nu}^{EM} &=& \frac{1}{2} \sum_{S_i,S_f}
\langle P_f | \hat J_\mu^{EM}| P_i \rangle^\dagger
\langle P_f | \hat J_\nu^{EM} | P_i \rangle \, , \\
W_{\mu\nu}^{I} &=& \frac{1}{2} \sum_{S_i,S_f} \left[
\langle P_f | \hat J_\mu^{EM} | P_i \rangle^\dagger
\langle P_f | \hat J_\nu^{NC} | P_i \rangle
+ h.c. \right] \, . \nonumber
\end{eqnarray}
%%%%
\subsection{Invariant amplitudes and isospin structure}
The hadronic matrix elements have the structure
\begin{equation}
\label{stru}
{\cal M} = \epsilon^\mu J_\mu \, ,
\end{equation}
where $\epsilon^\mu$ is an abbreviation for the leptonic
matrix element. These matrix elements may be decomposed
in isospace according to
\begin{equation}
\label{iso}
{\cal M} = \underbrace{\chi_{f}^{\dagger} \tau_\pi
\chi^{\ }_{i} }_{=: I^0} {\cal M}^0
+ \underbrace{\chi_{f}^{\dagger} \frac{1}{2} \{ \tau_\pi , \tau_3 \}
\chi^{\ }_{i} }_{=: I^+} {\cal M}^+
+ \underbrace{\chi_{f}^{\dagger} \frac{1}{2} [ \tau_\pi , \tau_3 ]
\chi^{\ }_{i} }_{=: I^-} {\cal M}^- \, ,
\end{equation}
with $\chi^{\ }_{f}$ and $\chi^{\ }_{i}$
the isospinors of the nucleon in the final and initial state,
respectively, and $\tau_\pi$ the isospin matrix characterizing
the produced pion. Furthermore,
the hadronic transition currents are decomposed into invariant
amplitudes \cite{Ad68}
\begin{eqnarray}
\label{ia}
% \langle P_f | \hat V_\mu^{(\pm,0)}| P_i \rangle &=&
V_\mu^{(\pm, 0)} &=& \sum_{j = 1}^6 V_{j}^{(\pm, 0)}(\nu,\nu_B,q^2)
\bar u(\vec{P_f}) M_\mu^j u(\vec{P_i}) \, , \\
% \langle P_f | \hat A_\mu^{(\pm,0)}| P_i \rangle &=&
A_{\mu}^{(\pm, 0)} &=& \sum_{j = 1}^8 A_{j}^{(\pm, 0)}(\nu,\nu_B,q^2)
\bar u(\vec{P_f}) N_\mu^j u(\vec{P_i}) \nonumber \, .
\end{eqnarray}
The amplitudes $V_{j}^{(\pm, 0)},A_{j}^{(\pm, 0)}$
depend on the three independent variables
$\nu,\nu_B$, and $q^2$, and the superscript
$(\pm, 0)$ refers to the isospin decomposition (\ref{iso}).
A reasonable choice for the vector $(M_\mu^j)$ and axial vector
$(N_\mu^j)$ operators is \cite{Ad68}:
\begin{equation}
\begin{array}{l@{\qquad}l}
M_\mu^1 = \frac{i}{2} \gamma_5 ( \gamma_\mu \dida{q} -
\dida{q} \gamma_\mu ) & \zeta_1^V = 1 \\
M_\mu^2 = - 2 i \gamma_5 ( P_\mu \; k^\pi \cdot q
- P \cdot q \; k^\pi_\mu ) & \zeta_2^V = 1 \\
M_\mu^3 = i \gamma_5 ( \gamma_\mu \; k^\pi \cdot q -
\dida{q} k^\pi_\mu ) & \zeta_3^V = -1 \\
M_\mu^4 = 2 i \gamma_5 [ (  \gamma_\mu \, P \cdot q -
\dida{q} P_\mu) -\frac{m}{2}(\gamma_\mu \dida{q} -
\dida{q} \gamma_\mu ) ] &  \zeta_4^V = 1 \\
M_\mu^5 =  - i \gamma_5 ( q_\mu \; k^\pi \cdot q -
q^2 k^\pi_\mu ) & \zeta_5^V = -1 \\
M_\mu^6 = i \gamma_5 ( q_\mu \dida{q} -
q^2 \gamma_\mu ) & \zeta_6^V = -1
\label{ia_v}
\end{array}
\end{equation}
and
\begin{equation}
\label{ia_a}
\begin{array}{l@{\qquad}l}
N_\mu^1 = \frac{i}{2} ( \gamma_\mu \dida{k^\pi} -
\dida{k^\pi} \gamma_\mu ) & \zeta_1^A = -1 \\
N_\mu^2 = -2 i P_\mu & \zeta_2^A = -1 \\
N_\mu^3 = -i k^\pi_\mu & \zeta_3^A = 1 \\
N_\mu^4 = -i m \gamma_\mu & \zeta_4^A = -1 \\
N_\mu^5 = 2 i \dida{q} P_\mu & \zeta_5^A = 1 \\
N_\mu^6 = i \dida{q} k^\pi_\mu & \zeta_6^A = -1 \\
N_\mu^7 = -i {q}_\mu & \zeta_7^A = 1 \\
N_\mu^8 = i \dida{q} {q}_\mu & \zeta_8^A = -1. \\
\end{array}
\end{equation}
The vector current operators are explicitly
gauge invariant by construction,
$q \cdot M^j \equiv 0 \quad \forall j = 1 \ldots 6$.
The constants $\zeta_j^V$ and $\zeta_j^A$ specify the
behavior of the invariant amplitudes under the crossing
transformation $\nu \rightarrow -\nu$\ \cite{Ad68},
\begin{eqnarray}
\label{cross}
V_{j}^{(\pm, 0)}(\nu,\nu_B,q^2) &=& (\pm, +) \zeta_j^V
V_{j}^{(\pm, 0)}(-\nu,\nu_B,q^2) \, , \\
A_{j}^{(\pm, 0)}(\nu,\nu_B,q^2) &=& (\pm, +) \zeta_j^A
A_{j}^{(\pm, 0)}(-\nu,\nu_B,q^2) \nonumber \, .
\end{eqnarray}
%%%%
\subsection{Multipole decomposition}
The vector and axial currents of (\ref{ia}) may be decomposed into a
multipole series following the work of Adler \cite{Ad68}.
The leading $S$-wave contributions to the currents are
\begin{eqnarray}
\label{mulzer}
\epsilon \cdot V &=& \frac{4 \pi i W}{m} \chi_f^{\dagger}
\big\{ \frac{\vec{\sigma}
\cdot \hat{q}}{\mid \vec{q}\mid \omega} (\epsilon_0 \, Q^2
+ \omega \; \epsilon \cdot q) L_{0+} - \vec{\sigma} \cdot
\vec{\epsilon}_T E_{0+} + \ldots \big\} \chi_{i} \; , \\
\epsilon \cdot A &=& \frac{4 \pi W}{m} \chi_{f}^{\dagger}
\big\{ \epsilon_0 \, {\cal L}_{0+} + i \vec{\sigma}
\cdot (\hat{q} \times \vec{\epsilon}) {\cal M}_{0+}
+ \epsilon \cdot q \;
{\cal H}_{0+} + \ldots \big\} \chi_i \; .\nonumber
\end{eqnarray}
In these equations all variables have to be expressed in the
{\it cm} frame, in particular the components of the 4-momentum transfer,
$q = (\omega, \vec{q})$, the polarization vector of the virtual photon,
$\epsilon = (\epsilon_0, \vec{\epsilon})$ and the nucleon
spin $\vec{\sigma}$. Furthermore, $Q^2 = - q^2$ and $\vec{\epsilon}_T$
is the polarization vector transverse to the direction of
the virtual photon.
The ellipses denote $P$-waves and higher multipoles. Since the
polarization vector $\epsilon_{\mu}$ is proportional to the transition
current of the electron, $j_{\mu}$ or $j_{\mu}^5$, the four-product
$\epsilon \cdot q$ vanishes exactly for the (conserved) vector current.
However, it can also be safely neglected in the case of the axial vector,
because the divergence of the axial current is proportional
to the mass of the electron.
The threshold values of the $S$-wave multipoles have been predicted by
general principles following the arguments of low energy theorems (LET).
Since these theorems assume Lorentz and gauge invariance and the PCAC
relation, they should be obeyed by our model, too.
However, there have been
recently reported large modifications to LET due to loop corrections.
Concerning the vector current these corrections are particular
large for neutral pion photoproduction at threshold \cite{Me91},
but do not play an important role for the inclusive cross section,
which is dominated by charged pion production.
However, it is interesting that two of
the multipoles for the axial vector have contributions containing the
$\pi N$ $\sigma$-term. According to \cite{Me94} these are
\begin{eqnarray}
\label{mul}
{\cal L}_{0+}^{(+)} &=& \frac{1}{3\pi m_{\pi} f_{\pi}} \left[ \sigma
(q^2 - m_{\pi}^2) - \frac{1}{4} \sigma (0) \right]
- \frac{a^+ f_\pi}{m_\pi} + O (m_{\pi})\, , \\
{\cal H}_{0+}^{(+)} &=& \frac{a^+ f_{\pi}}{q^2 - m_{\pi}^2}
+ \frac{\sigma (0) - \sigma (q^2 - m_{\pi}^2)}{12 \pi f_{\pi}
(q^2 - m_{\pi}^2)} + O (m_{\pi})\, , \nonumber
\end{eqnarray}
where $f_{\pi} = 93$ MeV is the pion decay constant and $a^+$ the isospin
even $S$-wave $\pi N$-scattering length. As we see from
(\ref{mulzer}) and the above considerations,
the multipole ${\cal H}_{0+}$ does not contribute in  the
limit of a vanishing lepton mass. The longitudinal multipole
${\cal L}_{0+}$, however, contributes and its threshold
value is dominated by the $\sigma$-term.
Unfortunately, it appears in combination with the vector coupling of
the $Z^0$ at the vertex of the electron, i.e. this interesting term is
suppressed by a factor $(4 \sin^2 \theta_W -1)$.
%%%%
\subsection{Explicit structure of the tensors}
In this subsection the explicit structure of the
Lorentz tensors will be given.
The electromagnetic lepton tensor has the familiar form
\begin{eqnarray}
\eta_{\mu\nu}^{EM}
&=& \eta_{\mu\nu}^{(s)}  + h \eta_{\mu\nu}^{(a)} \; ,
\end{eqnarray} where $h$
denotes the helicity of the incoming electron. The antisymmetric
part can be omitted for unpolarized nucleons,
because the electromagnetic hadronic tensor is symmetric in this case.
The interference tensor for the lepton is
\begin{eqnarray}
\eta_{\mu\nu}^{I}
&=& \gve (\eta_{\mu\nu}^{(s)} +h \eta_{\mu\nu}^{(a)} )
+ \gae (\eta_{\mu\nu}^{(a)} + h \eta_{\mu\nu}^{(s)}) \\
&=& \eta_{\mu\nu}^{I,(s)} + \eta_{\mu\nu}^{I,(a)} , \nonumber
\end{eqnarray} where
$\gve$ and $\gae$ denote the weak neutral current couplings of the
electron
\begin{eqnarray}
\gve  &=& \frac{1}{4 \sin \theta_W \cos \theta_W}
(-1 + 4 \sin^2 \theta_W) \, , \\
\gae &=& \frac{1}{4 \sin \theta_W \cos \theta_W}
\, . \nonumber
\end{eqnarray}
Using current conservation, the electromagnetic tensor of
the nucleon has the Lorentz structure
\begin{eqnarray}
W_{\mu\nu}^{EM} &=& -g_{\mu\nu} W_1^{EM}
+ P^i_\mu P^i_\nu \frac{W_2^{EM}}{m^2}
+ k^\pi_\mu k^\pi_\nu \frac{W_3^{EM}}{m^2}
- \frac{1}{2} (  P^i_\mu k^\pi_\nu + P^i_\nu k^\pi_\mu )
\frac{1}{k_\pi \cdot q \; P_i \cdot q \; m^2} \\ & &
(q^2 m^2 W_1^{EM} + ( P_i \cdot q)^2 W_2^{EM}
\vphantom{\frac{1}{2}}
+ ( k_\pi \cdot q )^2 W_3^{EM} - q^4 W_4^{EM} ) \; .
\vphantom{\frac{1}{2}} \nonumber \end{eqnarray}
The symmetric part of the hadronic interference tensor
$W_{\mu\nu}^{I}$ has the same structure as $W_{\mu\nu}^{EM}$.
The corresponding antisymmetric part is
\begin{eqnarray}
\label{Wia}
W_{\mu\nu}^{I,(a)} &=&
-i \epsilon_{\mu\nu\alpha\beta} P_i^\alpha q^\beta \frac{W_5^{I}}{m^2}
-i \epsilon_{\mu\nu\alpha\beta} k_\pi^\alpha P_i^\beta
\frac{W_{6}^{I}}{m^2} % \\ & &
-i \epsilon_{\mu\nu\alpha\beta} k_\pi^\alpha q^\beta \frac{W_{7}^{I}}{m^2}
-i ( P^i_\mu k^\pi_\nu - P^i_\nu k^\pi_\mu ) \frac{W_{8}^I}{m^2}
\\ & &
-i ( P^i_\mu \epsilon_{\nu\alpha\beta\gamma}
    - P^i_\nu \epsilon_{\mu\alpha\beta\gamma} )
   k_\pi^\alpha q^\beta P_i^\gamma \frac{W_{9}^I}{m^4}
% \nonumber \\ & &
-i ( k^\pi_\mu \epsilon_{\nu\alpha\beta\gamma}
    - k^\pi_\nu \epsilon_{\mu\alpha\beta\gamma} )
   k_\pi^\alpha q^\beta P_i^\gamma \frac{W_{10}^I}{m^4} \, .
\nonumber
\end{eqnarray}
As the axial currents are not conserved, there are in principle
additional terms proportional to $q_\mu$ or $q_\nu$.
Since these terms vanish after contraction with the leptonic
tensor (\ref{lept}), they have been omitted in (\ref{Wia})
right away. The structure functions $W_j(\nu,\nu_B,q^2)$
can be expressed in terms of the invariant amplitudes (\ref{ia})
in a straightforward way. The result of this
calculation is given in Appendix \ref{strf}.
%%%%
\section{Model for the Hadronic Currents}
\label{mod}
In this section we will present the
phenomenological model that is used to calculate the
hadronic currents. It contains contributions of the Born terms
and a phenomenological description of the $\Delta(1232)$ resonance.
%%%%
\subsection{Nonresonant contributions}
\begin{figure}[htb]
\centerline{
\epsfxsize=8.0cm
\epsffile{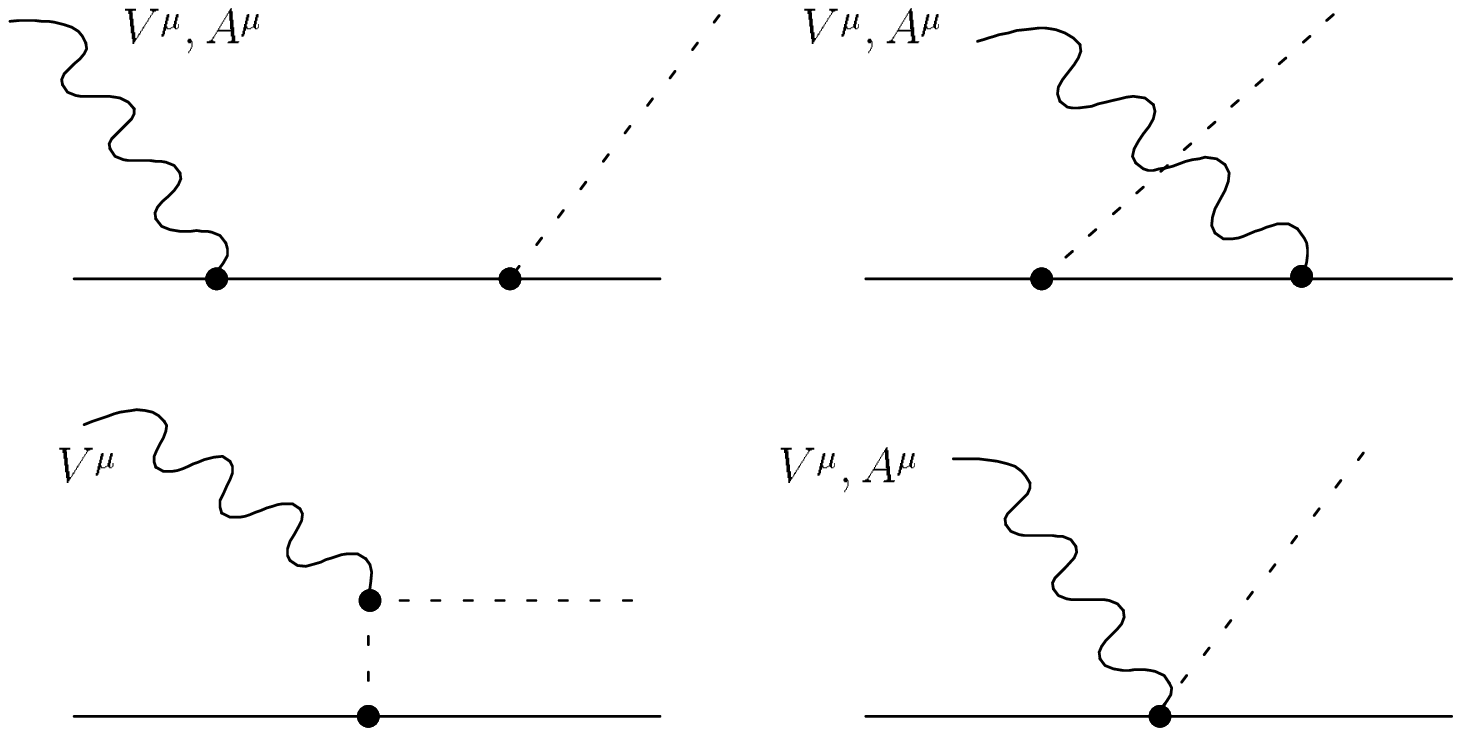}}
\caption{\label{fig3}
Feynman diagrams for nonresonant contributions.}
\end{figure}

The background of the Born terms contains both
vector current and axial vector current contributions.
Since we neglect the strangeness of the nucleon,
the weak vector current differs from the electromagnetic
current only by a coupling constant. Accordingly,
we have to calculate the Feynman diagrams of Fig. \ref{fig3}.
While all diagrams contribute to the vector current,
the pion pole diagram does not contribute to the axial current.
For the calculation of the Feynman diagrams we use the following
interaction Lagrangians
\begin{eqnarray}
\label{L}
{\cal L}_{\pi N N}^{PV} &=& \frac{f_{\pi N N}}{m_\pi} \bar{\psi}
\gamma_\mu \gamma_5 \vec{\tau} \psi \cdot \partial^\mu \vec{\phi}
\, , \\
{\cal L}_{V^\mu N N} &=& - e \, \bar{\psi} \frac{1}{2}
\bigg[ (\xi_V^{I=0} F_1^s + \xi_V^{I=1} \tau_3 F_1^v)
\gamma_\mu V^\mu
- (\xi_V^{I=0} \kappa_s F_2^s + \xi_V^{I=1} \kappa_v
\tau_3 F_2^v) \frac{\sigma_{\mu\nu}}{2 m} \partial^\nu V^\mu
\bigg] \psi \, , \nonumber \\
{\cal L}_{V^\mu \pi \pi} &=& -e\,\xi_V^{I=1} F_\pi
( \vec{\phi} \times \partial_\mu \vec{\phi}
)_3 V^\mu \, , \nonumber \\
{\cal L}_{V^\mu N N \pi} &=& e \, \xi_V^{I=1}
\frac{f_{\pi N N}}{g_a m_\pi} G_A \bar{\psi}
( \vec{\tau} \times \vec{\phi} )_3
\gamma_\mu \gamma_5 \psi \, V^\mu , \, \nonumber \\
{\cal L}_{A^\mu N N} &=& - e \, \xi_A^{I=1}
\bar{\psi} \left[ G_A \gamma_\mu A^\mu
+ G_P \frac{i \partial_\mu}{2 m} A^\mu
\right] \gamma_5 \frac{\tau_3}{2} \psi \, ,
\nonumber \\
{\cal L}_{A^\mu N N \pi} &=& e \, \xi_A^{I=1}
\frac{f_{\pi N N}}{ g_a m_\pi} \bar{\psi}
( \vec{\tau} \times \vec{\phi} )_3
\left( F_1^v \gamma_\mu A^\mu
- \kappa_v F_2^v \frac{ \sigma_{\mu\nu}}{2 m}
\partial^\nu A^\mu \right) \psi \, , \nonumber
\end{eqnarray}
with $\kappa_{(s,v)} = (\kappa_p \pm \kappa_n)$.
The isospin factors $\xi$ follow from the decomposition of
the corresponding quark current operators according to strong
isospin \cite{Mu94}. For the electromagnetic current these
factors are equal to unity, for the weak neutral current we have
\begin{eqnarray}
\xi_V^{I=1} &=& \frac{1}{2 \sin \theta_W \cos \theta_W}
( 1 - 2 \sin^2 \theta_W ) \, ,\\
\xi_V^{I=0} &=& -\frac{1}{2 \sin \theta_W \cos \theta_W} 2 \sin^2
\theta_W \, , \nonumber \\
\xi_A^{I=1} &=& -\frac{1}{2 \sin \theta_W \cos \theta_W} \, .
\nonumber
\end{eqnarray}
Note that there appears no isoscalar contribution to the axial
vector current, i.e. $\xi_A^{I=0} = 0$, because strange quarks
have been neglected.
The form factors $F_1$, $F_2$, $F_\pi$, $G_A$ and $G_P$
are functions of momentum transfer $q^2$. Since our calculation
will be performed at relatively small momentum transfer, we have
used simple dipole forms for the Sachs form factors, with the
assumption $G_A/g_a = F_\pi = F_1^v$. This insures gauge
invariance without additional gauge terms.
With standard methods and neglecting the lepton mass terms, we
obtain the invariant matrix elements
\begin{eqnarray}
\label{mat_b}
{\cal M}_s &=&  \xi_V^{I=1}
\frac{f_{\pi N N}}{m_\pi} \bar{u}(\vec{P_f})
\bigg[ \gamma_5 \dida{k_\pi}
\frac{\didag{P_i} + \dida{q} + m}{s - m^2}
\left( \tilde I_s^D \dida{\epsilon}
+ i \tilde I_s^P \frac{\sigma_{\mu\nu}}{2 m} q^\nu
\epsilon^\mu \right) \bigg] u(\vec{P_i})\, , \\
{\cal M}_u &=& \xi_V^{I=1}
\frac{f_{\pi N N}}{m_\pi} \bar{u}(\vec{P_f})
\bigg[ \left( \tilde I_u^D \dida{\epsilon}
+ i \tilde I_u^P \frac{\sigma_{\mu\nu}}{2 m} q^\nu
\epsilon^\mu \right)
\frac{\didag{P_i} - \dida{k_\pi} + m}{u - m^2}
\gamma_5 \dida{k_\pi} \bigg] u(\vec{P_i})\, , \nonumber \\
{\cal M}_t &=& \xi_V^{I=1}
\frac{f_{\pi N N}}{m_\pi}
\frac{\epsilon \cdot ( 2 k_\pi - q)}{t - m_\pi^2}
2 m I_t F_\pi
\bar{u}(\vec{P_f}) \gamma_5 u(\vec{P_i}) % \right]
\, , \nonumber \\
{\cal M}_{c} &=& \xi_V^{I=1}
\frac{f_{\pi N N}}{g_a m_\pi} I_t G_A
\bar{u}(\vec{P_f}) \dida{\epsilon} \gamma_5 u(\vec{P_i})
\, , \nonumber \\
{\cal M}_s^{5} &=& -\xi_A^{I=1}
\frac{f_{\pi N N}}{m_\pi} \bar{u}(\vec{P_f})
\left[ \dida{k_\pi} \frac{\didag{P_i} + \dida{q} - m}{s - m^2}
\tilde I_s^A \dida{\epsilon}
\right] u(\vec{P_i})\, , \nonumber \\
{\cal M}_u^{5} &=& -\xi_A^{I=1}
\frac{f_{\pi N N}}{m_\pi} \bar{u}(\vec{P_f}) \left[
\tilde I_u^A \dida{\epsilon}
\frac{\didag{P_i} - \dida{k_\pi} - m}{u - m^2}
\dida{k_\pi} \right] u(\vec{P_i})\, , \nonumber \\
{\cal M}_{c}^{5} &=& \xi_A^{I=1}
\frac{f_{\pi N N}}{g_a m_\pi} I_t \bar{u}(\vec{P_f}) \left[
\gamma_\mu F_1^v + i \frac{F_2^v}{2 m} \sigma_{\mu\nu} q^\nu
\right] \epsilon^\mu u(\vec{P_i})\, . \nonumber
\end{eqnarray}
Note that there do not appear any induced pseudoscalar terms,
because the lepton mass terms have been neglected.
The electromagnetic matrix element is the
sum of the first 4 terms in (\ref{mat_b}), corresponding to
s, u and t channel pole terms and the contact term(''c''). The matrix
element of the weak neutral current is the corresponding sum
of the combinations ${\cal M} +{\cal M}^5$.
The following abbreviations for the isospin matrix
elements have been used in (\ref{mat_b}):
\begin{eqnarray}
I_t &=& I^- ,  \\
\tilde I_{(s,u)}^D &=& \frac{1}{2} \left[
\frac{\xi_V^{I=0}}{\xi_V^{I=1}} F_1^s I^0
+ F_1^v ( I^+ \pm I^-) \right], \nonumber \\
\tilde I_{(s,u)}^P &=& \frac{1}{2} \left[
\frac{\xi_V^{I=0}}{\xi_V^{I=1}} F_2^s I^0
+ F_2^v ( I^+ \pm I^-) \right], \nonumber \\
\tilde I_{(s,u)}^A &=& \frac{1}{2} \left[
G_A ( I^+ \pm I^-) \right]
\, , \nonumber
\end{eqnarray}
with $I^{(\pm,0)}$ as defined in (\ref{iso}).
The invariant amplitudes obtained from (\ref{mat_b}) are
identical to the results of \cite{Ad68}.
They can be found in Appendix \ref{iam_b}.
%%%%
\subsection{Resonant contributions}
\begin{figure}[htb]
\centerline{
\epsfxsize=8.0cm
\epsffile{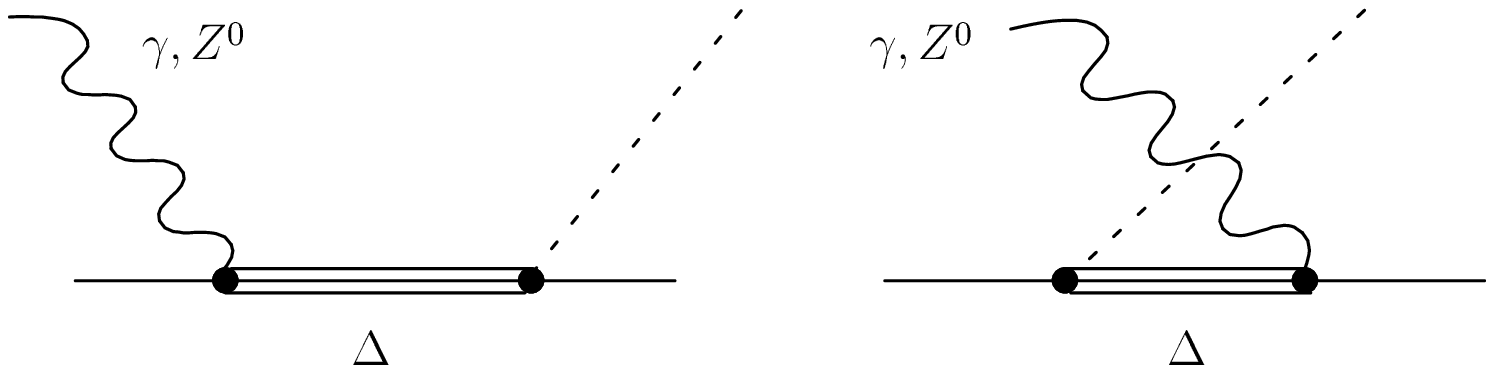}}
\caption{\label{fig4}
Feynman diagrams for resonant contributions.}
\end{figure}

In the kinematical region between
threshold and about 400 MeV excitation energy,
the dominant resonant contributions are due to the $\Delta(1232)$.
We treat the $\Delta$ as a Rarita-Schwinger field
and use the on-shell form of the propagator
modified by a phenomenological constant width,
\begin{eqnarray}
\label{d_prop}
\Delta_{\mu\nu} &=& \frac{\dida{p} + M}{p^2 -M^2 + i \Gamma M }
\left( g_{\mu\nu} -\frac{1}{3} \gamma_\mu \gamma_\nu
-\frac{2 p_\mu p_\nu}{3 M^2} +\frac{p_\mu \gamma_\nu -\gamma_\mu
p_\nu}{3 M} \right) \, .
\end{eqnarray}
$M = 1210 \mbox{ MeV}$ is obtained as
the real part of the $\Delta$ pole \cite{PDG94}.
The width $\Gamma = 85 \mbox{ MeV}$  is fitted to the
experimental data for pion electroproduction \cite{Co67,Ly67},
which is close to the imaginary part of the pole \cite{PDG94},
$\Gamma = 100 \mbox{ MeV}$ .
The Feynman diagrams to be calculated are shown in Fig. \ref{fig4}.
We use the
phenomenological $N \Delta$ transition currents \cite{Na82}
\begin{eqnarray}
% \langle \Delta(\vec{p'}) | \hat{J}_{\mu}^{EM} | N(\vec{p}) \rangle
{J}_{\mu}^{EM} &=& \bar{u}^{\lambda}(\vec{p'})
\left[ \left( \frac{C_3^{\gamma}}{m} \gamma^{\nu}
+ \frac{C_4^{\gamma}}{m^2} {p'}^{\nu} + \frac{C_5^{\gamma}}{m^2} p^{\nu}
\right) (g_{\lambda\mu} g_{\rho\nu} - g_{\lambda\rho} g_{\mu\nu})
q^{\rho} \gamma_5 \right] u(\vec{p})\, , \\
% & & \nonumber \\
% \langle \Delta(\vec{p'}) | \hat{J}_{\mu}^{NC} | N(\vec{p}) \rangle
{J}_{\mu}^{NC} &=& \bar{u}^{\lambda}(\vec{p'})
\left[ \left( \frac{C_{3V}^{Z}}{m} \gamma^{\nu}
+ \frac{C_{4V}^{Z}}{m^2} {p'}^{\nu} + \frac{C_{5V}^{Z}}{m^2} p^{\nu}
\right) \vphantom{\frac{1}{2}}
(g_{\lambda\mu} g_{\rho\nu} - g_{\lambda\rho} g_{\mu\nu})
q^{\rho} \gamma_5 + C_{6V}^Z g_{\lambda\mu} \gamma_5 \right.
\nonumber \\ & &+ \left. \left( \frac{C_{3A}^{Z}}{m}
\gamma^{\nu} + \frac{C_{4A}^{Z}}{m^2} {p'}^{\nu} \right)
(g_{\lambda\mu} g_{\rho\nu} - g_{\lambda\rho} g_{\mu\nu}) q^{\rho}
+ C_{5A}^Z g_{\lambda\mu} + \frac{C_{6A}^{Z}}{m^2} p_{\lambda}
q_{\mu} \right] u(\vec{p}) \, . \nonumber
\end{eqnarray}
The form factors $C_3$ to $C_6$
are taken from Nath et al.\cite{Na82},
with a $q^2$ dependence as in \cite{La88}.
The $\pi N \Delta$ vertex has the
effective form \cite{Ti78}
\begin{equation}
\label{piND}
\Lambda^{\mu}_{\pi N \Delta} = \frac{f_{\pi N \Delta}}{m_{\pi}}
k_\pi^{\mu}\, .
\end{equation}
With (\ref{d_prop})-(\ref{piND}) we obtain
the invariant matrix elements for s channel $\Delta$ production,
\begin{eqnarray}
{\cal M}^{EM}_{s \Delta} &=& \frac{f_{\pi N \Delta}}{m m_{\pi}}
\bar{u}(\vec{P_f}) k^{\pi}_{\sigma} \Delta^{\sigma \lambda}
\bigg[ C_3^{\gamma}( q_\lambda \gamma_\mu - g_{\lambda \mu} \dida{q})
+\frac{C_4^{\gamma}}{m}( q_\lambda(P_f + k_\pi)_{\mu}
\\ & &
-q \cdot ( P_f + k_\pi) g_{\lambda \mu} ) \bigg] \gamma_5
\epsilon^\mu u(\vec{P_i}) \, , \nonumber \\
{\cal M}^{NC}_{s \Delta} &=& \xi_V^{I=1}
{\cal M}^{EM}_{s \Delta} \vphantom{\frac{1}{2}}, \nonumber \\
{\cal M}^{5 NC}_{s \Delta} &=& -\xi_A^{I=1}
\frac{f_{\pi N \Delta}}{m m_{\pi}}
\bar{u}(\vec{P_f}) k^{\pi}_{\sigma} \Delta^{\sigma \lambda}
\bigg[ C_3^{A}( q_\lambda \gamma_\mu - g_{\lambda \mu} \dida{q})
+\frac{C_4^{A}}{m}( q_\lambda(P_f + k_\pi)_{\mu}
\nonumber \\ & &
-q \cdot ( P_f + k_\pi) g_{\lambda \mu} )
- C_5^{A} m g_{\lambda \mu}
\bigg] \epsilon^\mu u(\vec{P_i}) \, . \nonumber
\end{eqnarray}
The corresponding u channel amplitudes are constructed by using
the crossing properties of the invariant amplitudes
(\ref{cross}).
These invariant matrix elements can now be decomposed into
invariant amplitudes as in (\ref{ia}).
The result of this calculation is listed in Appendix \ref{iam}.
%%%%
\section{Results}
\label{res}
As a first test of the model we have calculated the inclusive
electromagnetic cross section. The experimental data for this
process \cite{Co67,Ly67} could be reproduced
within about 5\% over the whole energy region from threshold
to about 400 MeV excitation energy.

\begin{figure}[h]
\centerline{
\epsfxsize=18.0cm
\epsffile{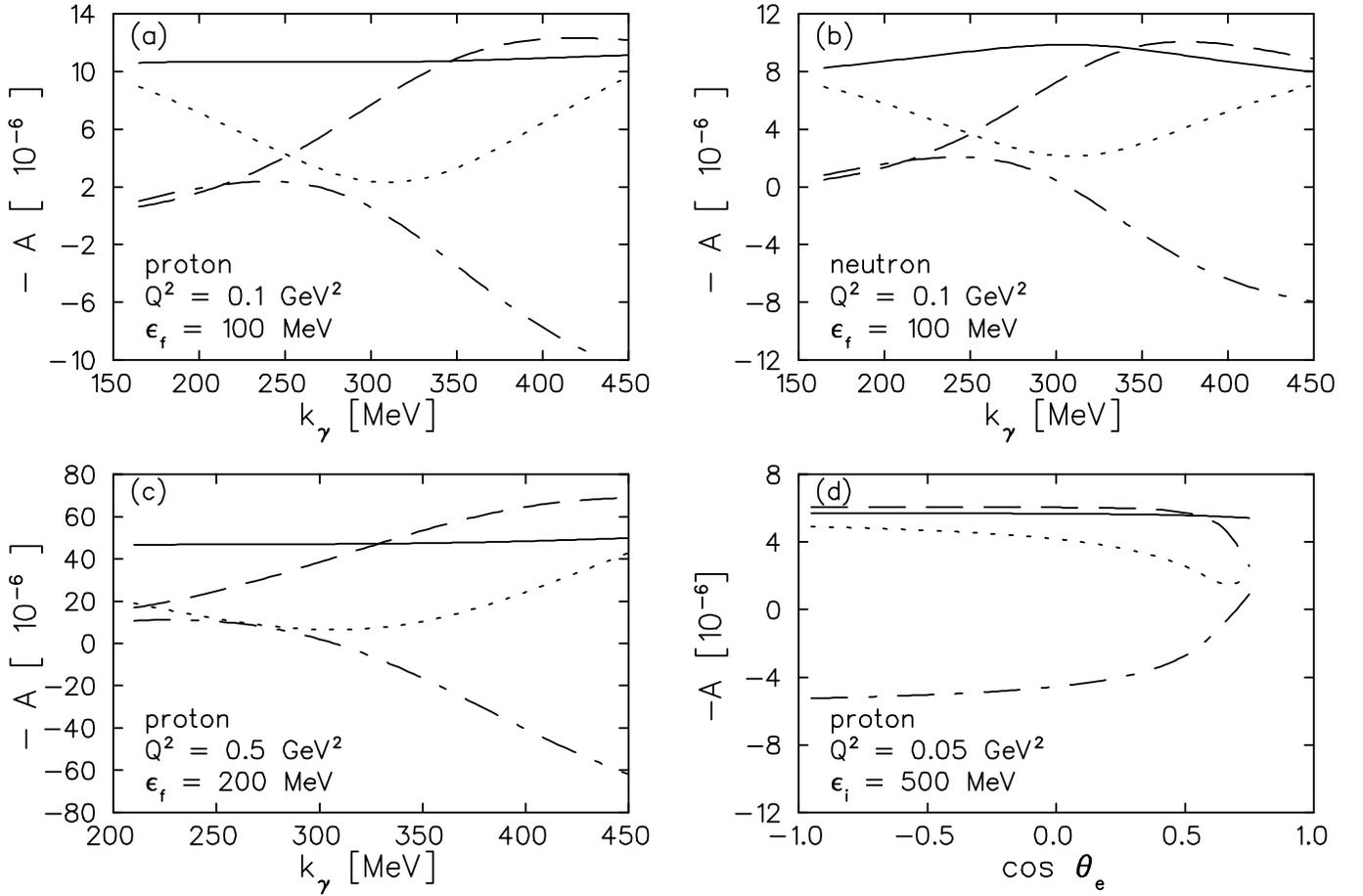}}
\caption{\label{fig5}
(a): Asymmetries for the proton as function of photon
equivalent energy at $Q^2 = 0.1 \mbox{ GeV}^2$.
The result of the full calculation (\solidl)
is compared to the contributions of background (\dotl),
resonance (\ldashl) and interference term (\dashdotl).
(b): Same as (a) for the neutron.
(c): Same as (a) for the proton at $Q^2 = 0.5 \mbox{ GeV}^2$.
(d): Same as (a) for the proton at $Q^2 = 0.05 \mbox{ GeV}^2$,
plotted as function of the electron scattering angle $\theta_e$.}
\end{figure}

Fig. \ref{fig5} shows the asymmetry for the proton
and the neutron in various kinematical situations.
The two upper figures compare the asymmetry for
proton and neutron as function of the photon equivalent energy,
\begin{equation}
k_{\gamma} = (W^2 - m^2)/2m \, ,
\end{equation}
the {\it lab} energy necessary to excite a hadronic system with
{\it cm} energy W. At small excitation energies the asymmetry is
essentially given by background contributions,
whereas the $\Delta$ resonance dominates at the higher
energies. While the individual contributions vary strongly,
the coherent sum is essentially constant. In particular,
the resonance structure of the inclusive cross sections
is essentially wiped out when the asymmetries are
calculated. As demonstrated in the two lower figures of
Fig. \ref{fig5}, this effect is quite independent of
the kinematics except for a trivial dependence on the
momentum transfer. For the proton this $Q^2$
dependence is well described by the simple estimate
\begin{equation}
\label{gr_a}
A \approx  -\frac{G_F Q^2}{e^2}  \approx -10^{-4} Q^2/\mbox{ GeV}^2 \; ,
\end{equation}
where $G_F$ is the Fermi coupling constant. In the case
of the neutron, the asymmetry is somewhat smaller,
at about 80\% of the proton case,
showing a slight enhancement in the resonance peak.

\begin{figure}[htb]
\centerline{
\epsfxsize=8.0cm
\epsffile{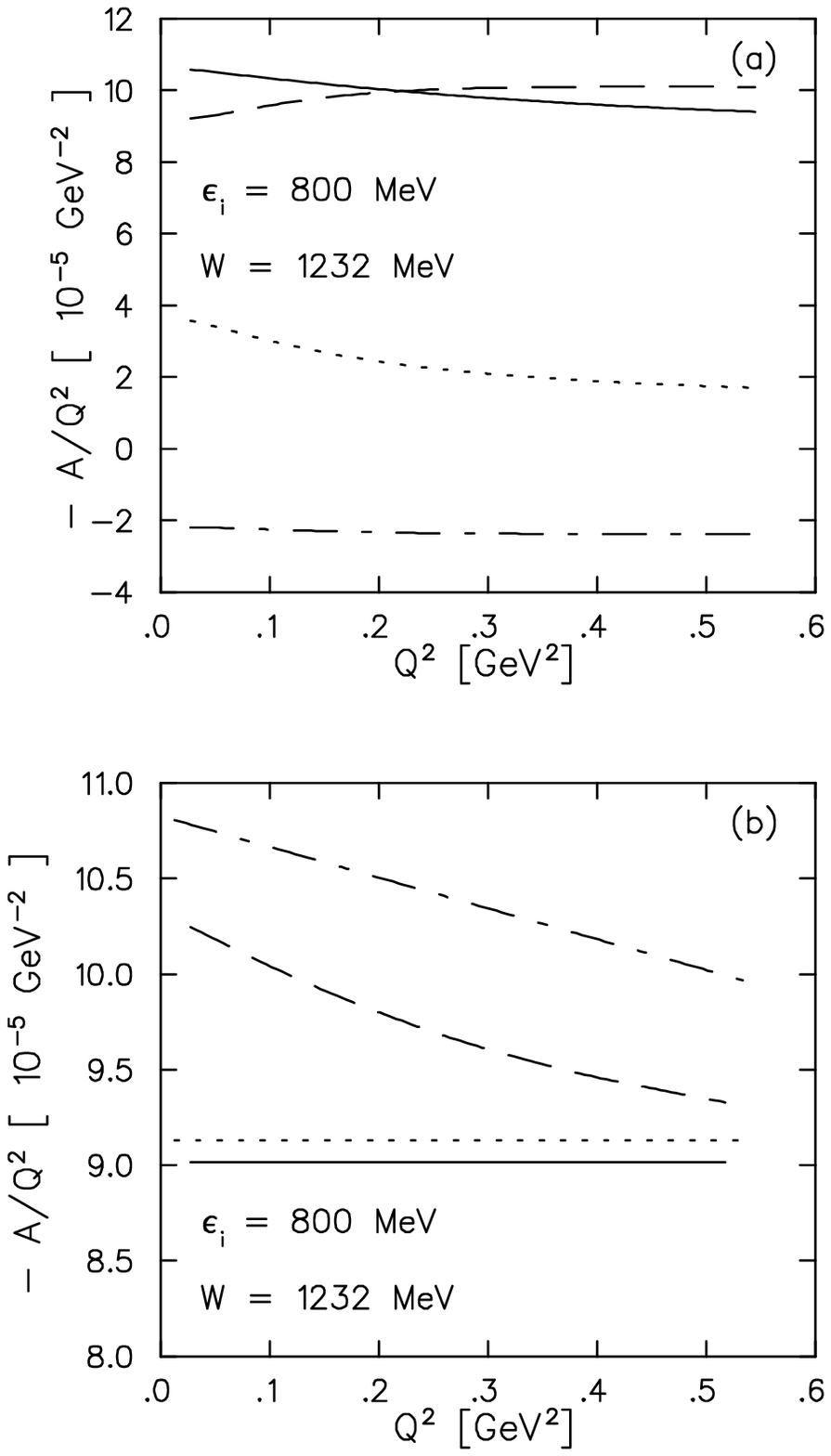}}
\caption{\label{fig6}
(a): Asymmetry for the proton in the resonance peak
as function of $Q^2$. For the notation of the curves
see Fig. \protect\ref{fig5}.
(b): Comparison of the asymmetry for the proton with earlier
calculations. Our result with $\Delta$ resonance only (\dashl)
should be compared with the work of Nath et al.\protect\cite{Na82}
(\dashdotl), and our result with $\Delta$ resonance only and hadronic
axial currents neglected (\solidl) should be compared with
the work of Cahn and Gilman \protect\cite{Ca78} (\dotl).}
\end{figure}

In Fig. \ref{fig6} the asymmetry is shown at an incident
electron energy of $\epsilon_i = 800 \mbox{ MeV}$ and at
$W = 1232 \mbox{ MeV}$, directly
in the resonance peak. In this kinematical situation,
the asymmetry is strongly dominated by the $\Delta$,
the background contributions cancel against the interference
terms (upper panel). Therefore, the neglect of
background terms as in the work of Nath et al.\cite{Na82}
is well justified. As is shown in the lower panel of this figure,
we are able to reproduce those results except for an overall
scaling factor of about 90 \%.
In the work of Cahn et al.\cite{Ca78} the vector
coupling between the $Z^0$ and the electron has been neglected.
This corresponds to $\sin^2 \theta_W = 1/4$
and produces a somewhat lower curve with a flat
distribution in $Q^2$.
We have also compared our calculation to the results of
Li et al. \cite{Li82} and find a reasonable agreement. However,
the calculation of Ishankuliev et al. \cite{Is80} disagrees
with all others by a factor $1/2$ seemingly due
to a wrong coupling constant.

Preliminary studies of the $\pi N$ $\sigma$-term (\ref{mul})
have shown only small
effects of this physically interesting quantity on the asymmetry.
Unfortunately, the coupling of the $Z^0$ to the axial current
of the hadron appears together with the vector coupling to
the electron which is strongly suppressed by the value of the
Weinberg angle. In addition, the $S$-wave multipole
${\cal L}_{0+}$ contributes to the asymmetry only by interference
with the electromagnetic $P$-waves. In the usual conventions
this contribution is proportional to $(2 M_{1+}
+ M_{1-}) {\cal L}_{0+}$. As a consequence, the
multipole ${\cal L}_{0+}$ is further suppressed near threshold.
%%%%
\section{Summary and Conclusions}
\label{conc}
The aim of this work has been to investigate parity violating (PV)
contributions to pion electroproduction. Only PV effects due to
interference between $\gamma$ and $Z^0$ exchange have been considered.
PV effects in the strong interaction,
as discussed by Li et al.\cite{Li82}, have been neglected.
The appropriate observable to study these effects is the
asymmetry defined in (\ref{def_a}).
A simple phenomenological model with effective Lagrangian densities
has been constructed that allows for the calculation
of the asymmetry in the kinematical region from pion threshold to
$\Delta(1232)$ resonance. The model is fully relativistic,
fulfils the crossing symmetry and
is gauge invariant due to our simple choice of the form factors.
The invariant amplitudes for the nonresonant contributions
respect PCAC. The $\Delta$ resonance is treated as a
Rarita-Schwinger field with phenomenological transition currents.
For the propagator, the on-shell form
with the $\Delta$ pole of the $\pi N$ scattering matrix is used.
The nonresonant contributions are created by the
usual pseudovector $\pi N$ coupling and phenomenological transition
currents. The weak neutral vector
current is traced back to the electromagnetic one by decomposing
the quark currents according to strong isospin.
For the axial current, a contact term has been introduced in
addition to the $s$ and $u$ channel nucleon Born terms.
This contact term is necessary to fulfil the low energy theorem of
Adler \cite{Ad68} for the nonresonant contributions.

The  calculated asymmetry is nearly constant as function of excitation
energy. It increases linearly with the square of the
four-momentum transfer, $A \approx 10^{-4} q^2 /\mbox{GeV}^2$
for the proton, and has about 80\% of that
value for the neutron.
Previous calculations for the excitation of the $\Delta$
isobar could be reproduced reasonably well, while our results are at
variance with some of the earlier work on the background contributions.

In conclusion we find the following results:
\begin{itemize}
\item
The asymmetry grows linearly with the momentum transfer $Q^2$ and is
nearly independent of the excitation energy.
\item
The expected asymmetries for pion electroproduction are
comparable with the asymmetries found for elastic
electron scattering.
\item
Because of the value of the Weinberg angle, the vector
coupling of the $Z^0$ at the electron vertex is suppressed.
As a consequence, the asymmetry is
dominated by the hadronic vector current,
the contributions of the hadronic
axial currents being of the order of 10 - 20 \% only.
\item
A precision measurement of the asymmetries is potentially an
independent experiment to determine the important
$\pi N$ $\sigma$-term, which appears
in the $S$-wave multipole ${\cal L}_{0+}$ of the hadronic axial current.
Unfortunately, the $\sigma$-term will be difficult to measure,
because the hadronic axial current is suppressed and furthermore,
the multipole ${\cal L}_{0+}$ appears in the asymmetry only by its
interference with $P$-wave multipoles.
\end{itemize}
Finally, we would like to comment on some possible improvements of the
calculations:
\begin{itemize}
\item
The simple superposition of Born and resonance terms violates the Watson
theorem. It would be necessary to perform a multipole decomposition
and to unitarize at least the multipoles carrying the phase of the
$\Delta(3,3)$ resonance.
\item
We have neglected the contribution of the strange sea to the hadronic
transition currents. Taking account of such effects would require
additional, as yet undetermined form factors. However,
it is unlikely that pion production would be strongly modified
by strangeness degrees of freedom.
\item
Though we have not found large contributions of the $\sigma$-term to the
asymmetry, this matter deserves more systematical studies. Clearly, any
further independent measurement of this important quantity would be
invaluable.
\end{itemize}
%%%%%%%%%%%%%%%%%%%%%%% ANHANG %%%%%%%%%%%%%%%%%%%%%%%%%%%%%%%%%%%%%%%%
\appendix
%%%%%%%%%%%%%%%%%%%%%%%%%%%%%%%%%%%%%%%%%%%%%%%%%%%%%%%%%%%%%%%%%%%%%%%
\section{Structure Functions}
\label{strf}
The amplitudes $V_i$ for the electromagnetic current and
$\tilde{V_i},\tilde{A_i}$ for the weak neutral current have the
following isospin decomposition for Born terms and
$\Delta$ contributions:
\begin{eqnarray}
V_i &=& I^0 V_i^0 + I^+ ( V_i^+ + V_{i,\Delta}^+)
+ I^- (V_i^- + V_{i,\Delta}^-)\, , \\
\tilde V_i &=& I^0 \tilde V_i^0 + I^+ ( \tilde V_i^+ + \tilde V_{i,\Delta}^+)
+ I^- (\tilde V_i^- + \tilde V_{i,\Delta}^-)\, , \nonumber \\
\tilde A_i &=& I^+ ( \tilde A_i^+ + \tilde A_{i,\Delta}^+)
+ I^- (\tilde A_i^- + \tilde A_{i,\Delta}^-)\, . \nonumber
\end{eqnarray}
%%%%
\\ {\bf Structure functions for the interference tensor} \\ \\
%%%%
{\bf Symmetric part} \\
\begin{eqnarray}
\label{W_sym}
W_1^{I} &=& \Re ( V_1^* \tilde V_1 )
\left( 4 m^2 ( \nu^2 - \nu_B^2) + m_\pi^2 q^2 \right)
+ \Re ( V_3^* \tilde V_3 ) 4 m^2 \nu_B^2 \left( 4 m^2 -
(4 m \nu_B + m_\pi^2 + q^2 ) \right) \nonumber \\
&+& \Re ( V_4^* \tilde V_4 )
4 m^2 \left( m_\pi^2 q^2 - 4 m^2 \nu_B^2
-\nu^2 (4 m \nu_B + m_\pi^2 + q^2 ) \right) \nonumber \\
&+& \Re ( V_6^* \tilde V_6 )
q^4 \left( 4 m^2 - (4 m \nu_B + m_\pi^2 + q^2 )
\right) - ( \Re( V_1^* \tilde V_3 ) + \Re( V_3^* \tilde V_1 ))
8 m^3 \nu \nu_B \nonumber \\
&+& (\Re( V_1^* \tilde V_4 ) + ( V_4^* \tilde V_1 ))
2 m \left( 4 m^2 \nu_B^2 - m_\pi^2 q^2 \right)
- (\Re( V_1^* \tilde V_6 ) + \Re( V_6^* \tilde V_1 ))
4 m^2 q^2 \nu \nonumber \\
&+& (\Re( V_3^* \tilde V_4 ) + \Re( V_4^* \tilde V_3 ))
4 m^2 \nu \nu_B (4 m \nu_B + m_\pi^2 + q^2 ) \nonumber \\
&+& (\Re( V_3^* \tilde V_6 ) + \Re( V_6^* \tilde V_3 ))
2 m \nu_B q^2
\left( 4 m^2- (4 m\nu_B +m_\pi^2 + q^2 )\right) \nonumber \\
&+& (\Re( V_4^* \tilde V_6 ) + \Re( V_6^* \tilde V_4 ))
2 m \nu q^2 (4 m \nu_B + m_\pi^2 + q^2 )\nonumber \\
\nonumber \\
%%%%
W_2^{I} &=& - 4 m^2 \big[ \Re (V_1^* \tilde V_1) q^2
+ \Re ( V_2^* \tilde V_2 ) 4 m^2 \nu_B^2
(4 m \nu_B + m_\pi^2 + q^2 ) \nonumber \\
&-& \Re ( V_3^* \tilde V_3 )
4 m^2 \nu_B^2 + \Re ( V_4^* \tilde V_4 )
\left( 4 m^2 \nu_B^2 - m_\pi^2 q^2 \right)
- \Re ( V_6^* \tilde V_6 ) q^4 \nonumber \\
&+& (\Re( V_1^* \tilde V_2 ) +\Re( V_2^* \tilde V_1 ))
2 m \nu_B ( q^2 + 2 m \nu_B)
- (\Re( V_3^* \tilde V_6 ) +\Re( V_6^* \tilde V_3 ))
2 m \nu_B q^2 \big] \nonumber \\
\nonumber \\
%%%%
W_3^{I} &=& - m^2 \big[ \Re ( V_2^* \tilde V_2 )
( \nu_B -\nu)^2 ( 4 m \nu_B + m_\pi^2 + q^2) \nonumber\\
&-& \Re ( V_3^* \tilde V_3 )
( 4 m^2 ( \nu_B -\nu)^2 +q^2( 4 m \nu_B + m_\pi^2 -4 m^2))
\nonumber \\ &+& \Re ( V_4^* \tilde V_4 )
( 4 m^2 ( \nu_B -\nu)^2 -q^2( 4 m \nu + m_\pi^2 4 m^2))
+\Re ( V_5^* \tilde V_5 )
q^4 (4 m \nu_B + m_\pi^2 + q^2) \nonumber \\
&+& (\Re( V_1^* \tilde V_2 )+ \Re( V_2^* \tilde V_1 ))
2 m ( 2 m (\nu_B -\nu)^2 + q^2(\nu_B -\nu)) \nonumber \\
&-& (\Re( V_1^* \tilde V_3 )+ \Re( V_3^* \tilde V_1 ))
2 m q^2 +(\Re( V_1^* \tilde V_4 )+ \Re( V_4^* \tilde V_1 ))
2 m q^2 \nonumber \\
&-& (\Re( V_1^* \tilde V_5 )+ \Re( V_5^* \tilde V_1 ))
q^2 ( 2 m (\nu_B -\nu) +q^2)
-(\Re( V_2^* \tilde V_3 )+ \Re( V_3^* \tilde V_2 ))
4 m^2 q^2 (\nu_B -\nu) \nonumber \\
&-& (\Re( V_2^* \tilde V_5 )+ \Re( V_5^* \tilde V_2 ))
2 m q^2 (\nu_B -\nu)
(4 m \nu_B + m_\pi^2 + q^2) \nonumber \\
&+& (\Re( V_2^* \tilde V_6 )+ \Re( V_6^* \tilde V_2 ))
4 m^2 q^2 (\nu_B -\nu)
+(\Re( V_3^* \tilde V_4 )+ \Re( V_4^* \tilde V_3 ))
q^2( 2 m(\nu_B +\nu) +m_\pi^2) \nonumber \\
&+& (\Re( V_3^* \tilde V_5 )+ \Re( V_5^* \tilde V_3 ))
2 m q^4 -(\Re( V_3^* \tilde V_6 )+ \Re( V_6^* \tilde V_3 ))
q^2( 2 m(\nu_B -\nu) +q^2) \nonumber \\
&+& (\Re( V_4^* \tilde V_6 )+ \Re( V_6^* \tilde V_4 ))
q^2( 2 m(\nu_B -\nu) +q^2)
-(\Re( V_5^* \tilde V_6 )+ \Re( V_6^* \tilde V_5 ))
2 m q^4 \big] \nonumber \\
\nonumber \\
%%%%
W_4^{I} &=& m^2 \big[ \Re ( V_1^* \tilde V_1 )
(4 m \nu + m_\pi^2 - q^2 ) - \Re ( V_2^* \tilde V_2 )
4 m^2 \nu_B^2 (4 m \nu_B + m_\pi^2 + q^2 ) \nonumber \\
&+& \Re ( V_4^* \tilde V_4 )
(4 m^2(m_\pi^2 -(\nu +\nu_B)^2) + m_\pi^2 ( q^2 - 4 m \nu))
- \Re ( V_5^* \tilde V_5 )
4 m^2 \nu_B^2 (4 m \nu_B + m_\pi^2 + q^2 ) \nonumber \\
&+&  \Re ( V_6^* \tilde V_6 )
(4 m^2 ( q^2 + \nu^2 - \nu_B^2)
- q^2 ( 4 m ( \nu +\nu_B ) + m_\pi^2 ) ) \nonumber \\
&-& (\Re( V_1^* \tilde V_2 ) + \Re( V_2^* \tilde V_1 ))
2 m \nu_B ( 2 m ( \nu_B - \nu) +q^2 )
-(\Re( V_1^* \tilde V_3 ) + \Re( V_3^* \tilde V_1 ))
4 m^2 \nu_B \nonumber \\
&-& (\Re( V_1^* \tilde V_4 ) + \Re( V_4^* \tilde V_1 ))
2 m m_\pi^2
-(\Re( V_1^* \tilde V_5 ) + \Re( V_5^* \tilde V_1 ))
2 m \nu_B (2 m (\nu_B - \nu) +q^2 ) \nonumber \\
&-& (\Re( V_1^* \tilde V_6 ) + \Re( V_6^* \tilde V_1 ))
4 m^2 \nu
-(\Re( V_2^* \tilde V_3 ) + \Re( V_3^* \tilde V_2 ))
8 m^3 \nu_B^2 \nonumber \\
&-& (\Re( V_2^* \tilde V_5 ) + \Re( V_5^* \tilde V_2 ))
4 m^2 \nu_B^2 (4 m \nu_B + m_\pi^2 + q^2 )
+(\Re( V_2^* \tilde V_6 ) + \Re( V_6^* \tilde V_2 ))
8 m^3 \nu_B^2 \nonumber \\
&+& (\Re( V_3^* \tilde V_4 ) + \Re( V_4^* \tilde V_3 ))
2 m \nu_B (2 m (\nu_B - \nu) +m_\pi^2 )
-(\Re( V_3^* \tilde V_5 ) + \Re( V_5^* \tilde V_3 ))
8 m^3 \nu_B^2 \nonumber \\
&+& (\Re( V_3^* \tilde V_6 ) + \Re( V_6^* \tilde V_3 ))
2 m \nu_B (4 m^2 -2 m (\nu_B - \nu) -m_\pi^2 ) \nonumber \\
&+& (\Re( V_4^* \tilde V_6 ) + \Re( V_6^* \tilde V_4 ))
2 m ( (2 m \nu_B + q^2)(\nu+ \nu_B) +\nu m_\pi^2 ) \nonumber \\
&+& (\Re( V_5^* \tilde V_6 ) + \Re( V_6^* \tilde V_5 ))
8 m^3 \nu_B^2 \big] \nonumber
\end{eqnarray}
%%%%
{\bf Antisymmetric part}
\begin{eqnarray}
\label{W_an}
W_5^I &=& - 2 m^2 \big[ \Re ( V_1^* \tilde A_1 )
m_\pi^2 + \Re ( V_3^* \tilde A_4 ) 2 m^2 \nu_B
- \Re ( V_4^* \tilde A_1 ) 2 m m_\pi^2
- \Re ( V_4^* \tilde A_4 ) 2 m^2 \nu \nonumber \\
&+& \Re ( V_6^* \tilde A_4 ) m q^2 \big] \nonumber \\
\nonumber \\
%%%%
W_{6}^I &=& 2 m^3 \big[ \Re ( V_1^* \tilde A_1 ) 2(\nu_B -\nu)
+ \Re ( V_3^* \tilde A_1 ) 4 m \nu_B
- \Re ( V_3^* \tilde A_4 ) 2 m \nu_B
- \Re ( V_4^* \tilde A_1 ) 4 m \nu_B \nonumber \\
&+& \Re ( V_4^* \tilde A_4 ) 2 m \nu
+ \Re ( V_6^* \tilde A_1 ) 2 q^2 - \Re ( V_6^* \tilde A_4 )
q^2 \big] \nonumber \\
\nonumber \\
%%%%
W_{7}^I &=& 2 m^3 \big[ \Re ( V_1^* \tilde A_4 ) m
+ \Re ( V_3^* \tilde A_1 ) 2 m \nu_B
- \Re ( V_4^* \tilde A_1 ) 2 m \nu
- \Re ( V_4^* \tilde A_4 ) 2 m^2 \nonumber \\
&+& \Re ( V_6^* \tilde A_4 ) q^2 \big] \nonumber \\
\nonumber \\
%%%%
W_{8}^I &=& 0 \nonumber \\  \nonumber \\
%%%%
W_{9}^I &=& 2 m^4 \big[ \Re ( V_1^* \tilde A_1 )
- \Re ( V_1^* \tilde A_2 )
+ \Re ( V_2^* \tilde A_1 ) 2 m \nu_B
+ \Re ( V_3^* \tilde A_5 ) 2 m \nu_B \nonumber \\
&+& \Re ( V_4^* \tilde A_2 ) 2 m
- \Re ( V_4^* \tilde A_4 ) m
- \Re ( V_4^* \tilde A_5 ) 2 m \nu
+ \Re ( V_6^* \tilde A_5 ) q^2 \big] \nonumber \\
\nonumber \\
%%%%
W_{10}^I &=& - m^4 \big[ - \Re ( V_1^* \tilde A_1 )
- \Re ( V_1^* \tilde A_2 )
+ \Re ( V_1^* \tilde A_3 )
+ \Re ( V_2^* \tilde A_1 ) 2 m (\nu_B -\nu) \nonumber \\
&-& \Re ( V_3^* \tilde A_1 ) 2 m
+ \Re ( V_3^* \tilde A_4 ) m
+ \Re ( V_3^* \tilde A_5 ) 2 m \nu_B
- \Re ( V_3^* \tilde A_6 ) 2 m \nu_B \nonumber \\
&+& \Re ( V_4^* \tilde A_1 ) 4 m
+ \Re ( V_4^* \tilde A_2 ) 2 m
- \Re ( V_4^* \tilde A_3 ) 2 m
- \Re ( V_4^* \tilde A_4 ) m \nonumber \\
&-& \Re ( V_4^* \tilde A_5 ) 2 m \nu
+ \Re ( V_4^* \tilde A_6 ) 2 m \nu
- \Re ( V_5^* \tilde A_1 ) q^2
+ \Re ( V_6^* \tilde A_5 ) q^2 \nonumber \\
&-& \Re ( V_6^* \tilde A_6 ) q^2 \big] \nonumber
\end{eqnarray}
%%%%
{\bf Structure functions for the electromagnetic tensor}\\ \\
The structure functions for the electromagnetic tensor follow
from (\ref{W_sym}) via the substitutions
\begin{eqnarray}
\Re ( V_k^* \tilde V_k ) &\rightarrow& | V_k |^2 \nonumber\, , \\
(\Re( V_l^* \tilde V_k )+ \Re( V_k^* \tilde V_l )) &\rightarrow&
2 \Re ( V_l^* V_k ) \, , \qquad
\forall l,k = 1 \ldots 6\, . \nonumber
\end{eqnarray}
%
%%%%%%%%%%%%%%%%%%%%%%%%%%%%%%%%%%%%%%%%%%%%%%%%%%%%%%%%%%%%%%%%%%%%
\section{Invariant Amplitudes for the Nonresonant Contributions}
\label{iam_b}
{\bf Vector currents}
\begin{eqnarray}
V_{1}^{(0,+)} &=& i \frac{f_{\pi NN}}{2 m_{\pi}}
\left(F_{1}^{(s,v)}\left(
\frac{1}{\nu -\nu_{B}} -\frac{1}{\nu +\nu_{B}}\right)
+\frac{F_{2}^{(s,v)}}{m}\right) \nonumber \\
%%%%
V_{2}^{(0,+)} &=& i \frac{f_{\pi NN}}{2 m_{\pi}}
\left(-\frac{F_{1}^{(s,v)}}{2 m\nu_{B}}
\left( \frac{1}{\nu -\nu_{B}} -\frac{1}{\nu +\nu_{B}}\right) \right)
\nonumber \\
%%%%
V_{3}^{(0,+)} &=& i \frac{f_{\pi NN}}{2 m_{\pi}}
\left(-\frac{F_{2}^{(s,v)}}{2 m}
\left( \frac{1}{\nu -\nu_{B}} +\frac{1}{\nu +\nu_{B}}\right) \right)
\nonumber \\
%%%%
V_{4}^{(0,+)} &=& i \frac{f_{\pi NN}}{2 m_{\pi}}
\left(-\frac{F_{2}^{(s,v)}}{2 m}
\left( \frac{1}{\nu -\nu_{B}} -\frac{1}{\nu +\nu_{B}}\right) \right)
\nonumber \\
%%%%
V_{5}^{(0,+)} &=& V_{6}^{(0,+)} = 0 \nonumber \\ \nonumber \\
%%%%
V_{1}^{(-)} &=& i \frac{f_{\pi NN}}{2 m_{\pi}}
\left(F_{1}^{v}\left( \frac{1}{\nu -\nu_{B}} +\frac{1}{\nu +\nu_{B}}
\right) \right) \nonumber \\
%%%%
V_{2}^{(-)} &=& i \frac{f_{\pi NN}}{2 m_{\pi}}
\left(-\frac{F_{1}^{v}}{2 m\nu_{B}}
\left( \frac{1}{\nu -\nu_{B}} +\frac{1}{\nu +\nu_{B}}\right) \right)
\nonumber \\
%%%%
V_{3}^{(-)} &=& i \frac{f_{\pi NN}}{2 m_{\pi}}
\left(-\frac{F_{2}^{v}}{2 m}
\left( \frac{1}{\nu -\nu_{B}} -\frac{1}{\nu +\nu_{B}}\right) \right)
\nonumber \\
%%%%
V_{4}^{(-)} &=& i \frac{f_{\pi NN}}{2 m_{\pi}}
\left(-\frac{F_{2}^{v}}{2 m}
\left( \frac{1}{\nu -\nu_{B}} +\frac{1}{\nu +\nu_{B}}\right) \right)
\nonumber \\
%%%%
V_{5}^{(-)} &=& i \frac{f_{\pi NN}}{2 m_{\pi}}
\left( \frac{2 F_{1}^{v}}{\nu_{B}\left(q^{2} + 4 m \nu_{B}\right)}\right)
\nonumber \\
%%%%
V_{6}^{(-)} &=& 0 \nonumber
\end{eqnarray}
%%%%
The amplitudes $\tilde V_{j}^{(\pm,0)}$
for the neutral weak vector current follow from
\begin{eqnarray}
\tilde V_{j}^{(\pm)} &=& \xi_V^{I=1} V_{j}^{(\pm)},
\qquad j = 1, \ldots ,6 \, \\
\tilde V_{j}^{(0)} &=& \xi_V^{I=0} V_{j}^{(0)},
\qquad j = 1, \ldots ,6 \; . \nonumber
\end{eqnarray}
%%%%
\\ {\bf Axial currents}
%%%%
\begin{eqnarray}
A_{1}^{(+)} &=& i \frac{f_{\pi NN}}{2 m_{\pi}}
\left( G_A \left( \frac{1}{\nu -\nu_{B}}
+\frac{1}{\nu +\nu_{B}}\right) \right) \nonumber \\
%%%%
A_{3}^{(+)} &=& i \frac{f_{\pi NN}}{2 m_{\pi}}
\left( G_A \left( \frac{1}{\nu -\nu_{B}}
-\frac{1}{\nu +\nu_{B}}\right) \right) \nonumber \\
%%%%
A_{2}^{(+)} &=& A_{4}^{(+)} = A_{5}^{(+)}
= A_{6}^{(+)} = A_{7}^{(+)} = A_{8}^{(+)} = 0
\nonumber \\ \nonumber \\
%%%%
A_{1}^{(-)} &=& i \frac{f_{\pi NN}}{2 m_{\pi}}
\left( G_A \left( \frac{1}{\nu -\nu_{B}}
-\frac{1}{\nu +\nu_{B}}\right) + \frac{F_2^v}{m g_a}
\right) \nonumber \\
%%%%
A_{2}^{(-)} &=& i \frac{f_{\pi NN}}{2 m_{\pi}}
\left( - \frac{F_2^v}{m g_a} \right) \nonumber \\
%%%%
A_{3}^{(-)} &=& i \frac{f_{\pi NN}}{2 m_{\pi}}
\left( G_A \left( \frac{1}{\nu -\nu_{B}} +\frac{1}{\nu +\nu_{B}}\right)
\right) \nonumber \\
%%%%
A_{4}^{(-)} &=& i \frac{f_{\pi NN}}{2 m_{\pi}}
\left( -\frac{2}{m} G_A + ( F_1^v + F_2^v ) \frac{2}{m g_a}
\right) \nonumber \\
%%%%
A_{5}^{(-)} &=& A_{6}^{(-)} =
A_{7}^{(-)} = A_{8}^{(-)} = 0 \nonumber
\end{eqnarray}
%%%%
The amplitudes $\tilde A_{j}^{(\pm)}$
for the neutral weak axial current follow from
\begin{eqnarray}
\tilde A_{j}^{(\pm)} &=& \xi_A^{I=1} A_{j}^{(\pm)},
\qquad j = 1, \ldots ,8 \, .
\end{eqnarray}
%%%%%%%%%%%%%%%%%%%%%%%%%%%%%%%%%%%%%%%%%%%%%%%%%%%%%%%%%%%%%%%%%%%%%%%
\section{Invariant Amplitudes for the Resonant Contributions}
\label{iam}
In the following, we give the s channel contributions of
the resonance. The corresponding u channel contributions
may be obtained by crossing symmetry, see (\ref{cross}).
\\ \\ {\bf Vector currents}
\begin{eqnarray}
V_{1, \Delta}^s &=& i \frac{f_{\pi N \Delta}}{m m_\pi}
\frac{1}{ s - m_\Delta^2 + i \Gamma m_\Delta}
\bigg[  C_3^\gamma \bigg( -2 m \nu_B -\frac{2}{3}m_\pi^2
-\abd ( \abbb + \frac{m_\pi^2}{2} )\nonumber \\
& &( \abbb +m \abc +\frac{q^2}{2})
+\abe( ( \abbb - \frac{q^2}{2})\abc - m m_\pi^2) \bigg)
\nonumber \\
& &-\frac{C_4^\gamma}{m} \bigg(  -2 m^2 \nu_B
-\frac{1}{3}(\abc(\abbb +\frac{q^2}{2})
+m m_\pi^2) \nonumber \\ & &+ \abe(\abbb +\frac{m_\pi^2}{2})
(\abbb +m \abc +\frac{q^2}{2}) \bigg)\Bigg] \nonumber \\
%\nonumber \\
%%%%
V_{2, \Delta}^s &=& i \frac{f_{\pi N \Delta}}{m m_\pi}
\frac{1}{ s - m_\Delta^2 + i \Gamma m_\Delta}
\bigg[ C_3^\gamma \bigg( 1 + \abd \frac{1}{2}
( \abbb + \frac{m_\pi^2}{2}) \abg \nonumber \\
& &+ \frac{1}{2}\abe \abc \abg \bigg)
-\frac{C_4^\gamma}{m} \bigg( \frac{\abf}{2}
+\frac{1}{6} \abc \abg \nonumber \\
& &-\abe ( \abbb + \frac{m_\pi^2}{2} ) \frac{1}{2} \abg
\bigg) \bigg] \nonumber \\
%\nonumber \\
%%%%
V_{3, \Delta}^s &=& i \frac{f_{\pi N \Delta}}{m m_\pi}
\frac{1}{ s - m_\Delta^2 + i \Gamma m_\Delta}
\bigg[ C_3^\gamma \bigg(  \frac{\abc}{3}
-\abd ( \abbb +\frac{m_\pi^2}{2} ) \nonumber \\
& &\frac{1}{2} \abc
+ \abe ( \abb + \frac{m_\pi^2}{2} ) \bigg)
-\frac{C_4^\gamma}{m} \bigg( -(\nu m +\frac{q^2}{2})
-\frac{m_\pi^2}{6} \nonumber \\
& &+ \abe ( \abbb + \frac{m_\pi^2}{2})
\frac{1}{2} \abc \bigg) \bigg] \nonumber \\
%\nonumber \\
%%%%
V_{4, \Delta}^s &=& i \frac{f_{\pi N \Delta}}{m m_\pi}
\frac{1}{ s - m_\Delta^2 + i \Gamma m_\Delta}
\bigg[ C_3^\gamma \bigg( -\frac{2}{3} \abc
- \abd ( \abbb + \frac{m_\pi^2}{2}) \nonumber \\
& &\frac{1}{2}\abc -\abe \frac{1}{2} m_\pi^2 \bigg)
-\frac{C_4^\gamma}{m} \bigg( - m \nu_B -\frac{1}{6} m_\pi^2
+ \abe ( \abbb +\frac{m_\pi^2}{2} ) \nonumber \\
& &\frac{1}{2} \abc \bigg)
\bigg] \nonumber \\
%\nonumber \\
%%%%
V_{5, \Delta}^s &=& i \frac{f_{\pi N \Delta}}{m m_\pi}
\frac{1}{ s - m_\Delta^2 + i \Gamma m_\Delta}
\bigg[ C_3^\gamma \bigg(\frac{1}{2} (1 -\frac{\nu}{\nu_B})
(\abd (\abbb +\frac{m_\pi^2}{2}) \nonumber \\
& &+\abe\abc ) \bigg)
-\frac{C_4^\gamma}{m} \bigg( -\frac{1}{2} \abf
+ \frac{1}{2}(1 -\frac{\nu}{\nu_B}) \nonumber \\
& &(\frac{\abc}{3} -\abe(\abbb +\frac{m_\pi^2}{2}))
\bigg) \bigg] \nonumber \\
%\nonumber \\
%%%%
V_{6, \Delta}^s &=& i \frac{f_{\pi N \Delta}}{m m_\pi}
\frac{1}{ s - m_\Delta^2 + i \Gamma m_\Delta}  \bigg[
C_3^\gamma \bigg( \abd( \abbb +\frac{m_\pi^2}{2})\frac{1}{2}
\abc \nonumber \\
& &+\abe \frac{m_\pi^2}{2} \bigg)
-\frac{C_4^\gamma}{m} \bigg( m \nu_B + \frac{m_\pi^2}{6}
- \abe( \abbb +\frac{m_\pi^2}{2})\frac{1}{2} \abc
\bigg) \bigg] \nonumber
\end{eqnarray}
%%%%
The amplitudes $V_{j, \Delta}^{(\pm)}$
are calculated via
\begin{eqnarray}
V_{j, \Delta}^+ &=& \frac{2}{3} ( V_{j, \Delta}^s +
V_{j, \Delta}^u )\, , \\
V_{j, \Delta}^- &=& -\frac{1}{3} ( V_{j, \Delta}^s
- V_{j, \Delta}^u ) \, ,
\qquad j = 1, \ldots , 6 \; , \nonumber
\end{eqnarray}
and the $\tilde V_{j, \Delta}^{(\pm)}$
for the neutral weak vector current follow from
\begin{equation}
\tilde V_{j, \Delta}^{(\pm)} = \xi_V^{I=1} V_{j, \Delta}^{(\pm)}\, ,
\qquad j = 1, \ldots ,6 \; .
\end{equation}
\ \\
%%%%
{\bf Axial currents}
\begin{eqnarray}
A_{1, \Delta}^s &=& -i \frac{f_{\pi N \Delta}}{m m_\pi}
\frac{1}{ s - m_\Delta^2 + i \Gamma m_\Delta}
\bigg[ \frac{C_4^A}{m} \bigg( \frac{1}{2} \abc
( \abbb +\frac{q^2}{2}) \nonumber \\
& &-\abe ( \abbb + \frac{m_\pi^2}{2})(\abbb +\frac{q^2}{2}) \bigg)
- C_5^A m \bigg( -\frac{1}{2} \abc \nonumber \\
& &+ \abe ( \abbb +
\frac{m_\pi^2}{2}) \bigg) \bigg] \nonumber \\
%\nonumber \\
%%%%
A_{2, \Delta}^s &=& -i \frac{f_{\pi N \Delta}}{m m_\pi}
\frac{1}{ s - m_\Delta^2 + i \Gamma m_\Delta}
\bigg[ \frac{C_4^A}{m} \bigg( -m \nu_B \abc
- \frac{1}{3} \abc \abbb \nonumber \\
& &+ \abe ( \abbb + \frac{m_\pi^2}{2}) \abbb \bigg)
- C_5^A m \bigg(
-\abd ( \abbb + \frac{m_\pi^2}{2}) \nonumber \\
& &\frac{1}{2} \abc
-\frac{1}{2} \abe m_\pi^2 \bigg) \bigg] \nonumber \\
%\nonumber \\
%%%%
A_{3, \Delta}^s &=& -i \frac{f_{\pi N \Delta}}{m m_\pi}
\frac{1}{ s - m_\Delta^2 + i \Gamma m_\Delta}
\bigg[ \frac{C_4^A}{m} \bigg( -\abc ( m \nu
+\frac{q^2}{2}) +\frac{1}{6} \abc q^2 \nonumber \\
& &-\abe( \abbb +\frac{m_\pi^2}{2}) \frac{q^2}{2}
\bigg) - C_5^A m \bigg( \frac{2}{3} \abc -\abd
( \abbb + m_\pi^2) \nonumber \\
& &\frac{1}{2} \abc + \abe \abbb \bigg) \bigg] \nonumber \\
%\nonumber \\
%%%%
A_{4, \Delta}^s &=& -i \frac{f_{\pi N \Delta}}{m m_\pi}
\frac{1}{ s - m_\Delta^2 + i \Gamma m_\Delta} \bigg[
\frac{C_4^A}{m} \bigg( \frac{m_\pi^2}{3 m}(\abbb +\frac{q^2}{2})
-\abe( \abbb \nonumber \\
& & +\frac{m_\pi^2}{2}) (\abbb +\frac{q^2}{2}) \frac{\abc}{m}
\bigg) - C_5^A m \bigg( -\frac{m_\pi^2}{3 m}
+\abe \frac{\abc}{m} \nonumber \\
& &(\abbb +\frac{m_\pi^2}{2}) \bigg)
\bigg] \nonumber \\
%\nonumber \\
%%%%
A_{5, \Delta}^s &=& -i \frac{f_{\pi N \Delta}}{m m_\pi}
\frac{1}{ s - m_\Delta^2 + i \Gamma m_\Delta}
\bigg[ \frac{C_4^A}{m} \bigg( m \nu_B -\frac{1}{6}
( 2 m \abc - m_\pi^2) \nonumber \\
& &-\frac{1}{2} \abe \abf ( \abbb + \frac{m_\pi^2}{2}) \bigg)
- C_5^A m \bigg( \frac{1}{2} \abd (\abbb +\frac{m_\pi^2}{2})
\nonumber \\
& &+ \frac{1}{2} \abc \abe \bigg) \bigg] \nonumber \\
%\nonumber \\
%%%%
A_{6, \Delta}^s &=& -i \frac{f_{\pi N \Delta}}{m m_\pi}
\frac{1}{ s - m_\Delta^2 + i \Gamma m_\Delta}
\bigg[ \frac{C_4^A}{m} \bigg( \nu m +\frac{q^2}{2}
- \frac{1}{6} ( 2 m \abc - m_\pi^2) \nonumber \\
& &-\frac{1}{2} \abe \abf ( \abbb + \frac{m_\pi^2}{2}) \bigg)
- C_5^A m \bigg(  -1 +\frac{1}{2} \abd ( \abbb + m_\pi^2)
\nonumber \\
& &+ \frac{1}{2} \abc \abe \bigg) \bigg] \nonumber \\
%\nonumber \\
%%%%
A_{7, \Delta}^s &=& -i \frac{f_{\pi N \Delta}}{m m_\pi}
\frac{1}{ s - m_\Delta^2 + i \Gamma m_\Delta}
\bigg[ \frac{C_4^A}{m} \bigg( -m \abc \nu_B
- \frac{1}{6} \abb \abc \nonumber \\
& & +\abe( \abbb +\frac{m_\pi^2}{2}) \abbb \bigg)
- C_5^A m \bigg( -\abd ( \abbb +\frac{m_\pi^2}{2})
\nonumber \\ & &\frac{1}{2} \abc
- \abe \frac{m_\pi^2}{2} \bigg) \bigg] \nonumber \\
%\nonumber \\
%%%%
A_{8, \Delta}^s &=& -i \frac{f_{\pi N \Delta}}{m m_\pi}
\frac{1}{ s - m_\Delta^2 + i \Gamma m_\Delta}
\bigg[ \frac{C_4^A}{m} \bigg( m \nu_B - \frac{1}{6}
( 2 m \abc -m_\pi^2) \nonumber \\
& &-\abe( \abbb +\frac{m_\pi^2}{2}) \frac{1}{2} \abf \bigg)
- C_5^A m \bigg( \frac{1}{2}\abd(\abbb +
\frac{m_\pi^2}{2} ) \nonumber \\
& &+\frac{1}{2} \abc \abe \bigg) \bigg] \nonumber
\end{eqnarray}
%%%%
The amplitudes
$A_{j, \Delta}^{(\pm)}$
are calculated via
\begin{eqnarray}
A_{j, \Delta}^+ &=& \frac{2}{3}
( A_{j, \Delta}^s + A_{j, \Delta}^u )
\, , \\
A_{j, \Delta}^- &=& -\frac{1}{3}
( A_{j, \Delta}^s - A_{j, \Delta}^u ) \, ,
\qquad j = 1, \ldots , 8 \; , \nonumber
\end{eqnarray}
and the $\tilde A_{j, \Delta}^{(\pm)}$ for the
neutral weak axial current follow from
\begin{equation}
\tilde A_{j, \Delta}^{(\pm)} = \xi_A^{I=1} A_{j, \Delta}^{(\pm)}\, ,
\qquad j = 1, \ldots ,8 \; .
\end{equation}
%%%%%%%%%%%%%%%%%%%%%%%%%%%%%%%%%%%%%%%%%%%%%%%%%%%%%%%%%%%%%%%%%%%%
%
%%%%

%%%%%%%%%%%%%%%%%%%%%FIGURES%%%%%%%%%%%%%%%%%%%%%%%%%%%%%%%%%%%%%%%%%
\newpage
\centerline{
{\bf \LARGE Figures}
%{\bf \LARGE Figure captions}
}
%%%%
%%%%
%%%%
%%%%
%%%%
%%%%

\begin{thebibliography}{References}
\bibitem[1]{Pr78}
C.Y. Prescott et al.: Phys. Lett. {\bf B77}(1978)347;
ibid. {\bf B84}(1979)524
\bibitem[2]{He89}
W. Heil et al.: Nucl. Phys. {\bf B327}(1989)1
\bibitem[3]{So90}
P.A. Souder et al.: Phys. Rev. Lett. {\bf 65}(1990)694
\bibitem[4]{Be92}
D. Beck et al.: CEBAF {\bf PR-91-017}(1991, revised 1992)
\bibitem[5]{So91}
R. Souder et al.: CEBAF {\bf PR-91-010}(1991)
\bibitem[6]{Bei91}
E. Beise et al.: CEBAF {\bf PR-91-004}(1991)
\bibitem[7]{Ha93}
D. von Harrach et al.: Mainz proposal {\bf A4/1-93}(1993)
\bibitem[8]{MK89}
R. McKeown et al.: MIT/Bates proposal {\bf 89-06}(1989)
\bibitem[9]{Na82}
L.M. Nath, K. Schilcher, M. Kretzschmar:
Phys. Rev. {\bf D25}(1982)2300
\bibitem[10]{Jo80}
D.R.T. Jones, S.T. Petcov: Phys. Lett. {\bf B91}(1980)137
\bibitem[11]{Ca78}
R.N. Cahn, F.J.Gilman: Phys. Rev. {\bf D17}(1978)1313
\bibitem[12]{Is80}
D. Ishankuliev, M.Ya. Safin: Sov. J. Nucl. Phys. {\bf 31}(1980)512
\bibitem[13]{Li82}
S.P. Li, E.M. Henley, W.-Y.P. Hwang:
Ann. Phys.(NY) {\bf 143}(1982)372
\bibitem[14]{Mu94}
M.J. Musolf et al.: Phys. Rep. {\bf 239}(1994)1
\bibitem[15]{Ad68}
S.L. Adler: Ann. Phys.(NY) {\bf 50}(1968)189
\bibitem[16]{Co67}
A. Cone et al.: Phys. Rev. {\bf 156}(1967)1490;
ibid. {\bf 163}(1967)1854(E)
\bibitem[17]{Ly67}
H.L. Lynch et al.: Phys. Rev. {\bf 164}(1967)1635
\bibitem[18]{Re90}
P. Renton: Electroweak Interactions. Cambridge University Press 1990
\bibitem[19]{Me91}
V. Bernard, N. Kaiser, J. Gasser, U.-G. Mei\ss ner:
Phys. Lett. {\bf B268}(1991)291
\bibitem[20]{Me94}
V. Bernard, N. Kaiser, U.-G. Mei\ss ner:
Phys. Lett. {\bf B331}(1994)137
\bibitem[21]{PDG94}
Particle Data Group, Review of Particle Properties:
Phys. Rev. {\bf D50}(1994)1173
\bibitem[22]{La88}
J.M. Laget: Nucl. Phys. {\bf B481}(1988)765
\bibitem[23]{Ti78}
L. Tiator, H.J. Weber, D. Drechsel: Nucl. Phys. {\bf A306}
(1978)468
\end{thebibliography}
\end{document}